\documentclass[aps,prl,reprint,superscriptaddress,showpacs]{revtex4-2}

\usepackage[english]{babel}
\usepackage[latin2]{inputenc}

\usepackage{amsmath}
\usepackage{amssymb}
\usepackage{graphicx}
\usepackage{verbatim}
\usepackage{xcolor}
\usepackage{multirow}

\begin{document}

%\begin{comment}

\title{Stability of pattern formation in systems with dynamic source regions}

\author{M. Majka}
\email{maciej.majka@uj.edu.pl}
\affiliation{Institute of Theoretical Physics and Mark Kac Center for Complex Systems Research, Jagiellonian University, ul. prof. Stanis\l{}awa \L{}ojasiewicza 11, 30-348 Krak\'{o}w Poland}
\author{R.D.J.G. Ho}
\affiliation{Institute of Theoretical Physics and Mark Kac Center for Complex Systems Research, Jagiellonian University, ul. prof. Stanis\l{}awa \L{}ojasiewicza 11, 30-348 Krak\'{o}w Poland}
\affiliation{Current address: The Njord Centre, Department of Physics, University of Oslo, P.O. Box 1048, 0316 Oslo, Norway}
\author{M. Zagorski}
\email{marcin.zagorski@uj.edu.pl}
\affiliation{Institute of Theoretical Physics and Mark Kac Center for Complex Systems Research, Jagiellonian University, ul. prof. Stanis\l{}awa \L{}ojasiewicza 11, 30-348 Krak\'{o}w Poland}

\begin{abstract}
We explain the principles of gene expression pattern stabilization in systems of interacting, diffusible morphogens, with dynamically established source regions. Using a reaction-diffusion model with step-function production term, we identify the phase transition between low-precision indeterminate patterning and the phase in which a traveling, well-defined contact zone between two domains is formed. Our model analytically explains single- and two-gene domain dynamics and provides pattern stability conditions for all possible two-gene regulatory network motifs. 
\end{abstract}

\maketitle
%\end{comment}

%=====INTRODUCTION==========
Reaction-diffusion dynamics with threshold-enhanced production is encountered in many branches of physics, such as the study of combustion \cite{bib:combustion}, neural signaling \cite{bib:signal_old, bib:signal_propagation,bib:fitznagumo}, climate evolution \cite{bib:budyko1,bib:budyko2}, population dynamics \cite{bib:allee_effect,bib:KPP1,bib:Kessler_1998,bib:bacteria,bib:population}, chemical reactions and phase coexistence \cite{bib:phase_sep1,bib:phase_sep2}. Reaction diffusion dynamics is also the basic process that governs the spreading of morphogens across a developing tissue \cite{bib:Bollenbach_2005,bib:Kicheva_2007,bib:Gregor_2007,bib:Shvartsman_2012,bib:Muller_2013,bib:Stapornwongku_2021}. Target genes interpret morphogen signals to form gene expression patterns (GEPs). Many aspects of the patterning process have already been investigated, including the scaling of GEPs as an embryo grows \cite{bib:McHale_2006,bib:Ben-Zvi_2008,bib:Ben-Zvi_2010,bib:Umulis_2013,bib:Kicheva_2015,bib:Aguilar-Hidalgo_2018,bib:Mateus_2020}, the precision of domain boundary localization in emerging patterns \cite{bib:Zagorski_2017,bib:Petkova_2019,bib:Exelby_2021,bib:Jagere_2004,bib:Morishita_2011,bib:Hiornaka_2012,bib:Sokolowski_2012,bib:Tran_2018,bib:Jaeger_2020,bib:Tkacik_2021}, and the relations between structure and function of gene regulatory networks (GRNs) that drive pattern formation \cite{bib:Luscombe_2004,bib:Reeves_2006,bib:Alon_2007,bib:Cotterel_2010,bib:Burda_2011,bib:Lander_2011,bib:Balaskas_2012,bib:Martin_2016,bib:Jimenez_2017,bib:Verd_2017,bib:Verd_2019}. However, some important problems remain unaddressed. In particular, little is known about systems where diffusible gene products affect the size of their own source regions (domains). Such dynamic source regions could expand or shrink in unbounded manner, yet, these scenarios are mitigated by additional regulatory mechanisms. This is encountered in spinal cord development  \cite{bib:Ribes_2010,bib:Kuzmicz-Kowalska_2020,bib:Shinozuka_2021}, limb formation \cite{bib:Raspopovic_2014,bib:Marcon_2016,bib:Landge_2020,bib:Mateus_2020} and \textit{Drosophila} wing and eye development \cite{bib:Vuilleumier_2010,bib:Wartlick_2011,bib:Wartlick_2014}. In this letter we elucidate the general physical principles behind the GEP stabilization for one and two genes as well as any combination of regulatory interactions in the system. Whilst we use the language of genes, our analysis is not limited to biological context.

\begin{figure}
\includegraphics[width=1.0\columnwidth]{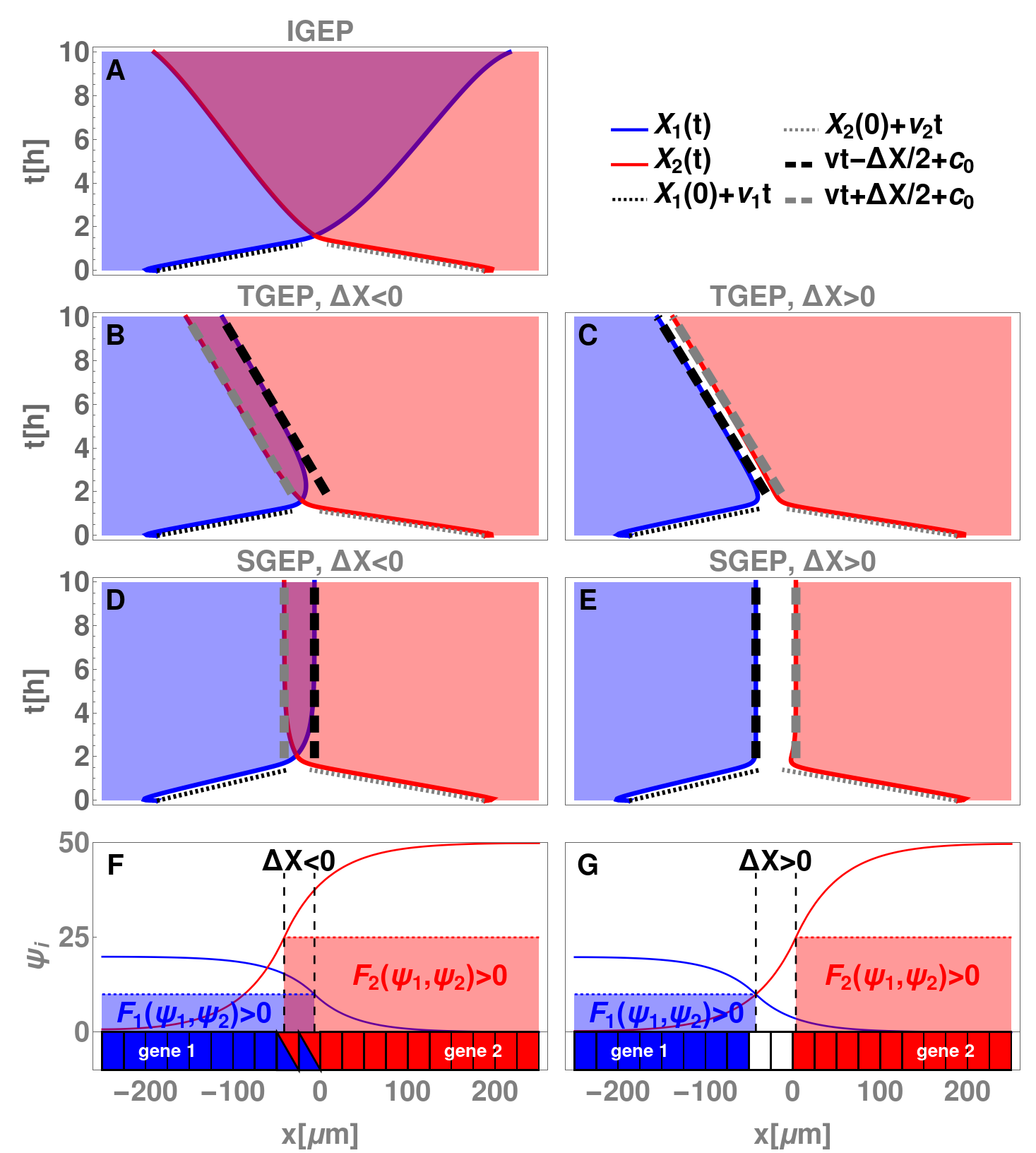}
\caption{Exemplary dynamics of GEPs generated by Eq.~\eqref{eq:main1}, for self-activating ($\epsilon_{ii}>0$) and cross-repressive ($\epsilon_{i\neq j}<0$) gene interactions. Other parameters are chosen in the biologically relevant range (see SM). Red and blue shading indicates the domains of active morphogen production, the positions of domain boundaries are found numerically ($X_i(t)$, red and blue solid line) and analytically (gray and black lines). A: IGEP, unbounded expansion of overlap. B and C: TGEP, fronts move to the left, for $t\lesssim 2$h (dotted lines) front velocities $v_i$ are found from Eq.~\eqref{eq:v2}. For $t\gtrsim 4$h (dashed lines) common velocity $v$ and width $\Delta X$ are found from Eq.~\eqref{eq:two_v}. $c_0$ is determined numerically to match the analytical and numerical results. D and E: SGEP, $v=0$, parameters satisfy stability conditions  \eqref{eq:stab_cond}. Contact zone is an overlap (B, D) or a gap (C, E). F and G: relation between morphogen concentration profiles, GEP and cellular interpretation of GEP.   \label{fig:2comp}}
\end{figure}

We focus our analysis on the contact zone between two gene expression domains, situated at the opposite sides of a system. The contact zone is either a gap between the domains or their partial overlap (see Fig.~\ref{fig:2comp}A-G). The gap corresponds to the stripe of undifferentiated cells. The overlap can be interpreted in two ways: either as the tissue co-expressing two specific target genes \cite{bib:Alaynick_2011} or as the imprecise boundary region between domains, where the actual cells in the developing tissue would commit to one of the two fates \cite{bib:Zagorski_2017,bib:Exelby_2021}. First, we provide exact classification of domain dynamics for one gene. Then, for two-gene systems, we characterize the phase transition between the phase with unbounded expansion of the overlap, leading to the \textit{indeterminate} GEP (IGEP, Fig.~\ref{fig:2comp}A) and the phase of \textit{traveling} GEP (TGEP). In the latter case, a stable, fixed-size contact zone is formed, though it can still travel as one entity, as the domains change size in a coordinated manner (Fig.~\ref{fig:2comp}B, C). The transition is controlled by the strength of gene-gene regulatory interactions. Among TGEPs, non-moving \textit{stable} GEPs (SGEPs) are identified, for which the drift velocity of contact zone is exactly zero (Fig.~\ref{fig:2comp}D, E). We identify the exact relations between system parameters that ensure the formation of SGEPs. Our results are mostly analytical, supported by numerics where necessary.

In the biological context, SGEP might be difficult to achieve as it requires specific combinations of system parameters. However, systems that can be mapped into the vicinity of SGEP in parameter space are guaranteed to form low-velocity TGEPs. Drifting GEPs were observed in \textit{Drosophila} (shifting of posterior gap gene domains) \cite{bib:Verd_2017,bib:Verd_2019}, in spinal cord development \cite{bib:Kicheva_2014,bib:Zagorski_2017,bib:Kicheva_2015} and limb formation \cite{bib:Raspopovic_2014,bib:Marcon_2016}. As development happens on finite time-scales, a slowly moving TGEPs and SGEPs might be similarly efficient in their biological role and both might be practically indistinguishable in experiments. 

A convenient model for investigating the GEP stability was first introduced to study the 4-gene domain size regulation mechanism in \textit{Drosophila} \cite{bib:vakulenko} and the stability of a single traveling domain subjected to extrinsic perturbations and intrinsic noise \cite{bib:rulands}. The model included reaction-diffusion equation with step-function production term. The approximation of interacting kinks, representing the domain boundaries, was utilized in both works to obtain the results. Here, we employ a similar model as in \cite{bib:vakulenko,bib:rulands}, but instead of using a moving kink approximation, we identify the conservation law, which allows for exact analytical treatment.

In this letter we will relate the diffusible gene products with morphogens and the gene expression domains with morphogen source regions. We consider two morphogens that undergo diffusion and degradation and are able to affect each other and their own production. The space-time concentration profile $\psi_i(x,t)$ ($\psi_i$, for brevity) of each morphogen obeys the equation:
\begin{equation}
\begin{split}
\partial_t \psi_i=&D_i \partial_{xx}\psi_i-\gamma_i\psi_i+H_i \theta\left(F_i(\psi_1,\psi_2) \right) \label{eq:main1}
\end{split}
\end{equation}
where we have diffusion constant $D_i$, degradation rate $\gamma_i$, production rate in activated state $H_i$, and Heaviside step function $\theta(\dots)$. Gene expression is often characterized by Hill-type kinetics with steep increase near the activation threshold \cite{bib:Gregor_2007b,bib:Sokolowski_2012,bib:Sokolowski_2015}, for which the Heaviside function is a generic approximation. The functions $F_i(\psi_1,\psi_2)$ are the activation conditions corresponding to the two-gene motif of the GRN. In linear approximation:
\begin{equation}
\begin{split}
&F_i(\psi_1,\psi_2)\simeq \epsilon_{ii}\psi_i+\epsilon_{ij}\psi_j-C_i \label{eq:activ_cond}
\end{split}
\end{equation}
Here $\epsilon_{ij}$ are the interaction coefficients, $\epsilon_{ij}>0$ indicates activation and $\epsilon_{ij}<0$ the inhibition of production. $C_i$ is the threshold for production, possibly affected by the external influence of GRN on the $i$-th node. Thus, we consider both $C_i>0$ (gene requires activation) and $C_i<0$ (gene is active by default).

The regions where $F_i(\psi_1,\psi_2)>0$ are identified as expression domains and constitute the GEP. The relation between morphogen concentration $\psi_i$ and underlying GEP is illustrated in Fig.~\ref{fig:2comp}F and G.

Four effective parameters characterize the system:
\begin{align}
\lambda_i=\sqrt{\frac{D_i}{\gamma_i}},&&\tilde \psi_i=\frac{H_i}{\gamma_i},&&S_i=\frac{2C_i}{\epsilon_{ii}\tilde \psi_i},&&\chi_i=\frac{\epsilon_{ij}\tilde \psi_j}{\epsilon_{ii}\tilde \psi_i} \label{eq:params}
\end{align}
$\lambda_i$ is the effective distance traveled by morphogen particle before degradation and it quantifies the range of interactions. $\tilde \psi_i$ is the equilibrium concentration level to which the $i$-th morphogen tends in the absence of cross-interactions ($\epsilon_{i\neq j}=0$) and diffusion. $S_i$ is the effective activation threshold and $\chi_i$ describes the relative strength of cross-to-auto interaction for the $i$-th gene. 

In order to study a single contact zone, we supply Eq.~\eqref{eq:main1} with the initial condition:
\begin{equation}
\psi_i(x,0)=A_i\theta(\sigma_i(x-X_i(0))) \label{eq:init_cond}
\end{equation}
where $X_i(t)$ is the position of the domain border (activation front), and $A_i$ is the initial concentration. $\sigma_i=\pm1$ indicates which side of the system is occupied by the $i$-th domain. With $A_i>C_i/\epsilon_{ii}$, these initial conditions ensure the formation of only one activation front per gene. We assume reflective boundary conditions and derive our results in the infinite system, $L\to+\infty$. This is a satisfying approximation to the in-the-bulk dynamics of finite-size systems, as long as $ L/2-|X_i(t)|>\lambda_i$. Limitations are discussed in the Supplemental Material (SM). 

%=====fig 1: two components========
%================

A remarkable property of \eqref{eq:main1} is that $\psi_i(x,t)$ can be found analytically without the prior knowledge of $X_i(t)$. The Green's function of eq. \eqref{eq:main1} reads:
\begin{equation}
G_i(x-x',t-t')=\frac{e^{-\gamma_i(t-t')-\frac{(x-x')^2}{4D_i(t-t')}}}{\sqrt{4 \pi D_i (t-t')}}
\end{equation}
Then, the concentration profile $\psi_i(x,t)$ reads:
\begin{equation}
\begin{split}
\psi_i(x,t)=&\int_{-\infty}^{+\infty}dx'G_i(x-x',t)\psi_i(x',0)+\\
& +\sigma_i H_i\int_0^tdt'\int^{\sigma_i\infty}_{X_i(t')}dx'G_i(x-x',t-t')
\end{split}\label{eq:sol}
\end{equation}
In order to obtain $X_i(t)$ one must solve the free boundary problem $F_i(\psi_1(X_i(t),t),\psi_2(X_i(t),t))=0$ or explicitly:
\begin{equation}
\begin{gathered}
\epsilon_{11}\psi_1(X_1(t),t)+\epsilon_{12}\psi_2(X_1(t),t)=C_1\\
\epsilon_{21}\psi_1(X_2(t),t)+\epsilon_{22}\psi_2(X_2(t),t)=C_2
\end{gathered}\label{eq:main2}
\end{equation}
By inserting \eqref{eq:sol} into \eqref{eq:main2}, one finds that $X_1(t)$ and $X_2(t)$ are defined by a system of non-linear integral equations. 

%=====fig 2: one component============
\begin{figure}
\centering
\includegraphics[width=0.95\columnwidth]{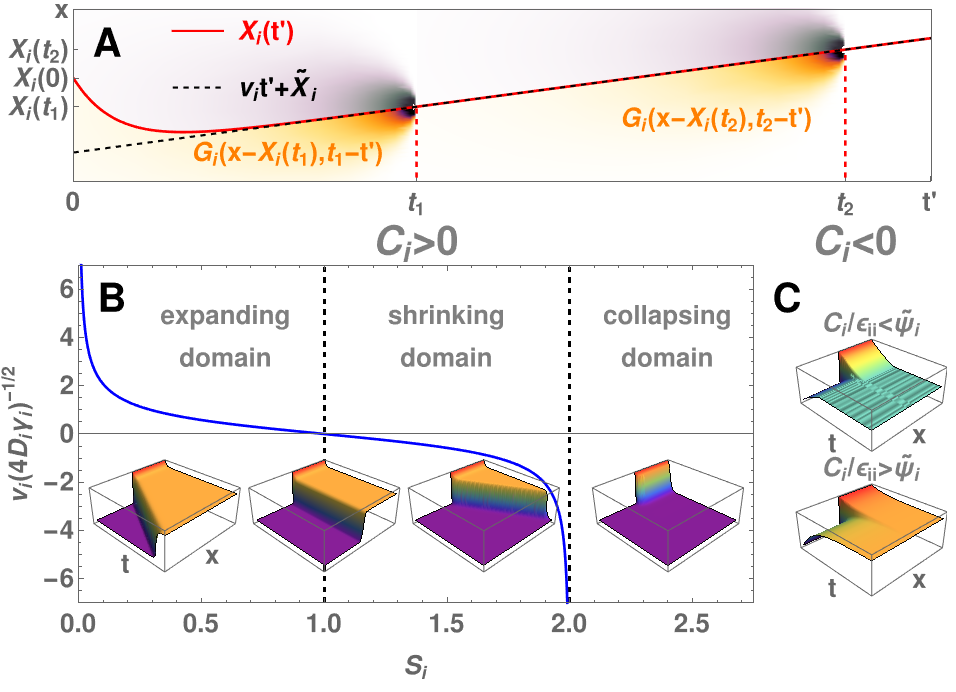}

\caption{A: Illustration of Eq. \eqref{eq:one_comp}, defining the evolution of domain boundary $X_i(t)$. Density maps in the background show the kernel $G_i(x-X_i(t_i),t-t_i)$ at two moments, $t_1$ and $t_2$. Color indicates the space-time area of integration in Eq. \eqref{eq:one_comp}. The integral of $G_i(x-X_i(t_i),t'-t_i)$ is asymptotically conserved on the line $X_i(t')=v_i t'+\tilde X_i$. B: Single domain dynamics of non-interacting gene for $\sigma_i=-1$ (domain in the region $(-\infty,X_i(t)]$), for $C_i>0$. $v_i$ is calculated from Eq. \eqref{eq:v2}. Insets show corresponding concentration profiles $\psi_i(x,t)$ in different regimes ($S_i=0.5$, $1.0$, $1.5$, $2.1$, from left to right). C: Single-gene dynamics for $C_i<0$ (spontaneous expression). For $C_i/\epsilon_{ii}>\tilde \psi_i$ concentration saturates at $\tilde \psi_i$ (lower), for $C_i/\epsilon_{ii}<\tilde \psi_i$ at $C_i/\epsilon_{ii}$ (upper). \label{fig:1comp}}
\end{figure}
%========================

Let us first consider the system without cross-interactions between target genes ($\epsilon_{i\neq j}=0$), so \eqref{eq:main2} reduces to $\epsilon_{ii}\psi_i(X_i(t),t)=C_i$. This means that the number of particles at the domain boundary is constant. For $t\gg \gamma_i^{-1}$ we can neglect the influence of initial conditions and multiply this equation by $(\epsilon_{ii}\tilde \psi_i)^{-1}$ to obtain:
\begin{equation}
\frac{S_i}{2}=\sigma_i\gamma_i\int_0^tdt'\int^{\sigma_i \infty}_{X_i(t')}dx' G_i(X_i(t)-x',t-t') \label{eq:one_comp}
\end{equation}
This shows that $X_i(t)$ must evolve in such way that the value of the space-time integral of $G_i(X_i(t)-x',t-t')$ is conserved. This integral represents the number of particles arriving at $X_i(t)$ from the activated region, as $G_i$ is also the transition probability. Graphical analysis, as in Fig.~\ref{fig:1comp}A, shows that this is satisfied for:
\begin{equation}
X_i(t)=v_it+\tilde X_i \label{eq:ansatz}
\end{equation}
where $\tilde X_i$ is a constant. In SM we also show that $\dot X_i(t)=v_i$ is the attractor of dynamics. Similar results were rigorously proven by Terman \cite{bib:terman}, but our method can be easily extended to the interacting case. The ansatz \eqref{eq:ansatz} allows explicit integrations in Eq. \eqref{eq:one_comp} (see SM), which, in the limit $t\to+\infty$, leads to the equation:
\begin{equation}
S_i=1+\sigma_i\frac{v_i}{\sqrt{4D_i\gamma_i+v_i^2}} \label{eq:v2} 
\end{equation}
In Fig. \ref{fig:1comp}B we show $v_i$ calculated from \eqref{eq:v2} (see SM), which indicates a few different regimes of dynamics. First, let us consider $C_i>0$, which means gene is inactive by default. For $0<S_i<1$ the expression domain expands with constant velocity, while for $1<S_i<2$ it shrinks. For $S_i=1$ it is long-term stable. $S_i>2$ and $S_i<0$ translate into $\epsilon_{ii}\tilde \psi_i<C_i$. This means the long-term concentration is too low to sustain activation and an activated domain must collapse in its entire volume over finite time.
 
 For $C_i<0$, gene is spontaneously expressed everywhere in undifferentiated tissue and travelling fronts do not form, see Fig.~\ref{fig:1comp}C. For $\epsilon_{ii}\tilde \psi_i>C_i$, the long-term concentration tends to $\tilde \psi_i$, but for $\epsilon_{ii}\tilde \psi_i\le C_i$, the system saturates at the highest expression level just before inactivation, which is $C_i/\epsilon_{ii}$. This requires $\epsilon_{ii}<0$.
 
Let us now consider the fully interacting case, that is $\epsilon_{ij}\neq 0$ for $i\neq j$. We can repeat the reasoning for the non-interacting case, though this time it is the sum of the auto-interaction integral and cross-interaction integral that has to be conserved. The ansatz \eqref{eq:ansatz} holds, but with the important change that both velocities must be the same, i.e. $v_1=v_2=v$ (see SM). This allows us to turn \eqref{eq:main2} into the algebraic problem:
\begin{equation}
\begin{split}
&S_i=\left(1+\sigma_iV_i(v)\right)+\chi_i\left[1+(-1)^i\sigma_j\textrm{sgn}(\Delta X)+\right.\\
&+\sigma_j e^{\frac{v\Delta X}{2D_j}\left( (-1)^{i+1}-\frac{\textrm{sgn}(\Delta X)}{V_j(v)}\right)}\left.\left(V_j(v)-(-1)^i \textrm{sgn}(\Delta X) \right)\right]
\end{split} \label{eq:two_v}
 \end{equation}
 where $V_i(v)=\frac{v}{\sqrt{4D_i\gamma_i+v^2}}$ and $\Delta X=\tilde X_2-\tilde X_1$ is the relative distance between the domain boundaries. We will now discuss the properties of \eqref{eq:two_v} using $\chi_i$ (the ratio of cross-to-auto interaction strength) as control parameters.
 
\begin{figure}
\includegraphics[width=\columnwidth]{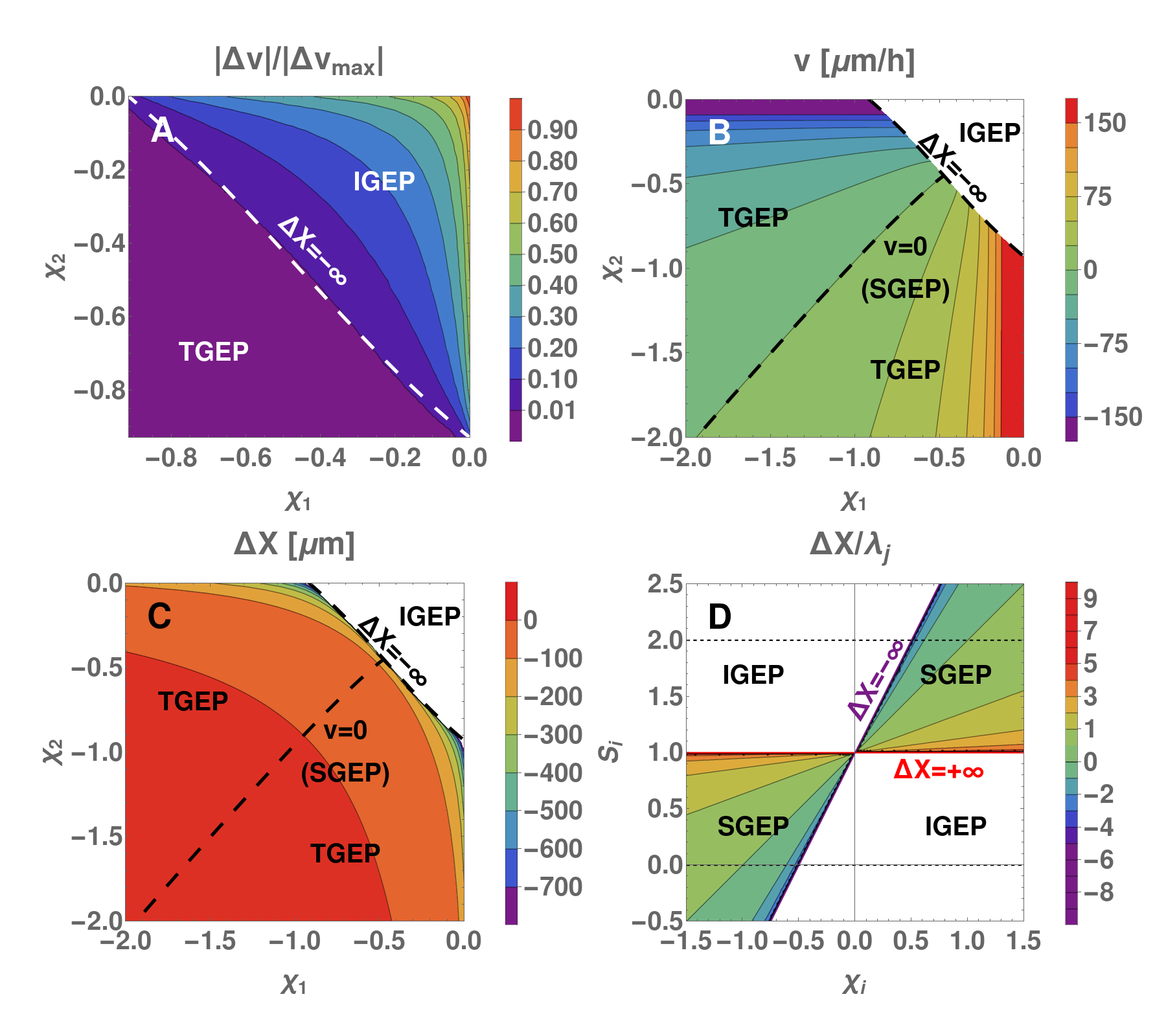}
\caption{A: Map of $|\Delta v|/|\Delta v_{max}|$ over the $(\chi_1,\chi_2)$ plane. $\Delta v_{max}$ is $\Delta v$ for $(\chi_1,\chi_2)=(0,0)$. $\Delta v$ is the order parameter of transition between IGEP and TGEP phase. Phase boundary coincides with $\Delta X=-\infty$ line (white, dashed). 3D representation of these data is included in SM. B and C: the maps of $v$ and $\Delta X$ over the $(\chi_1,\chi_2)$ plane in TGEP phase. The line $v=0$ indicate SGEPs. D: The phase diagram of SGEP, in the $(\chi_i,S_i)$ space. Colored region indicates where $\Delta X$ is defined. Stabilization requires that $\Delta X(\chi_1,S_1)=\Delta X(\chi_2,S_2)$.
 \label{fig:phase_trans}}
\end{figure}

First of all, Eq.~\eqref{eq:two_v} indicates the existence of a phase transition between the IGEP and  TGEP phases. For clearest presentation, let us consider two auto-activating and cross-inhibiting genes ($\chi_i<0$), whose expression domains spontaneously expand ($0<S_i<1$). The order parameter of transition reads: 
 \begin{equation}
 \Delta v=\lim_{t\to+\infty}(\dot X_2(t)-\dot X_1(t))
 \end{equation}
 which is the long-term difference of domain wall velocities. $\Delta v$ dependence on $(\chi_{1},\chi_2)$ is shown in Fig.~\ref{fig:phase_trans}A. In the IGEP phase $\Delta v\neq 0$, which means that two domains ever-increase their overlap and their boundaries eventually adopt two constant, but different velocities. The $t\to+\infty$ value of $\Delta X$ does not exists and 
 Eqs.~\eqref{eq:two_v} have no solution. We analytically estimated that the IGEP phase is no smaller than the square ${0>\chi_i>(S_i-1)/2}$ on the $(\chi_1,\chi_2)$ plane (see SM). However, simulations show that the boundary of IGEP phase coincides with the line $\Delta X=-\infty$ on the $(\chi_1,\chi_2)$ plane.

In the TGEP phase $\Delta v=0$ and Eqs.~\eqref{eq:two_v} can be solved for $v$ and $\Delta X$. In this case, upon meeting, the two growing domains establish a contact zone of width $\Delta X$. TGEP domains can still change their size by one expanding and the other shrinking, but the contact zone travels as one entity, preserving $\Delta X$ (see Figs.~\ref{fig:2comp}B, C). $\Delta X$ and $v$ as the functions of $(\chi_1,\chi_2)$ are shown in Fig.~\ref{fig:phase_trans}B, C. The IGEP-TGEP transition demonstrates that establishing a meaningful GEP requires certain minimal strengths of cross-interactions, below which no patterning is possible. Once TGEP phase is entered, further increase in interaction strengths drives $\Delta X$ from the complete overlap on the critical line ($\Delta X=-\infty$) to the emergence of the gap ($\Delta X\ge 0$). 

%======Fig 3: phase transition=============
%======================

Finally, SGEPs can be identified as TGEPs with $v=0$. The SGEPs occupy a line on the $(\chi_1,\chi_2)$ plane (Fig. \ref{fig:phase_trans}B, C). In this case Eqs.~\eqref{eq:two_v} are overdefined and ensuring that $\Delta X$ exists leads to the stability conditions. Let us introduce the auxiliary variable $R_i$, such that:
\begin{align}
R_i=(S_i-1)/\chi_i-1 \label{eq:R}
\end{align}

then, the stability conditions read:
\begin{subequations}
\begin{gather}
(1-|R_1|)^{\lambda_2}=(1-|R_2|)^{\lambda_1} \label{eq:cond_a}\\
\begin{aligned}
\sigma_1\textrm{sgn}(R_1)=-\sigma_2\textrm{sgn}(R_2),&&-1\le R_i\le 1
\end{aligned}
\end{gather}\label{eq:stab_cond}
\end{subequations}
When these dependencies are satisfied, two activated domains form SGEP with the distance between domain boundaries given by (for $i\neq j$):
\begin{align}
\Delta X=(-1)^j\sigma_i\textrm{sgn}(R_j)\lambda_i \ln (1-|R_j|) \label{eq:deltaX}
\end{align}

SGEPs are perfectly stable, i.e. they survive for $t\to+\infty$, but the condition \eqref{eq:cond_a} is restrictive and imposes strong constraints on the parameters. Nevertheless, SGEP indicates the center of low-velocity plateau in TGEP phase (see Fig. \ref{fig:phase_trans}B). The biologically relevant part of this plateau can be estimated to $\pm$24$\mu$m/h, based on the shift of gap gene domains in Drosophila \cite{bib:Jaeger_2004,bib:Verd_2017,bib:Markov_2009} (see SM). 

%==========Fig 4: stability regions==============
\begin{figure}
\includegraphics[width=\columnwidth]{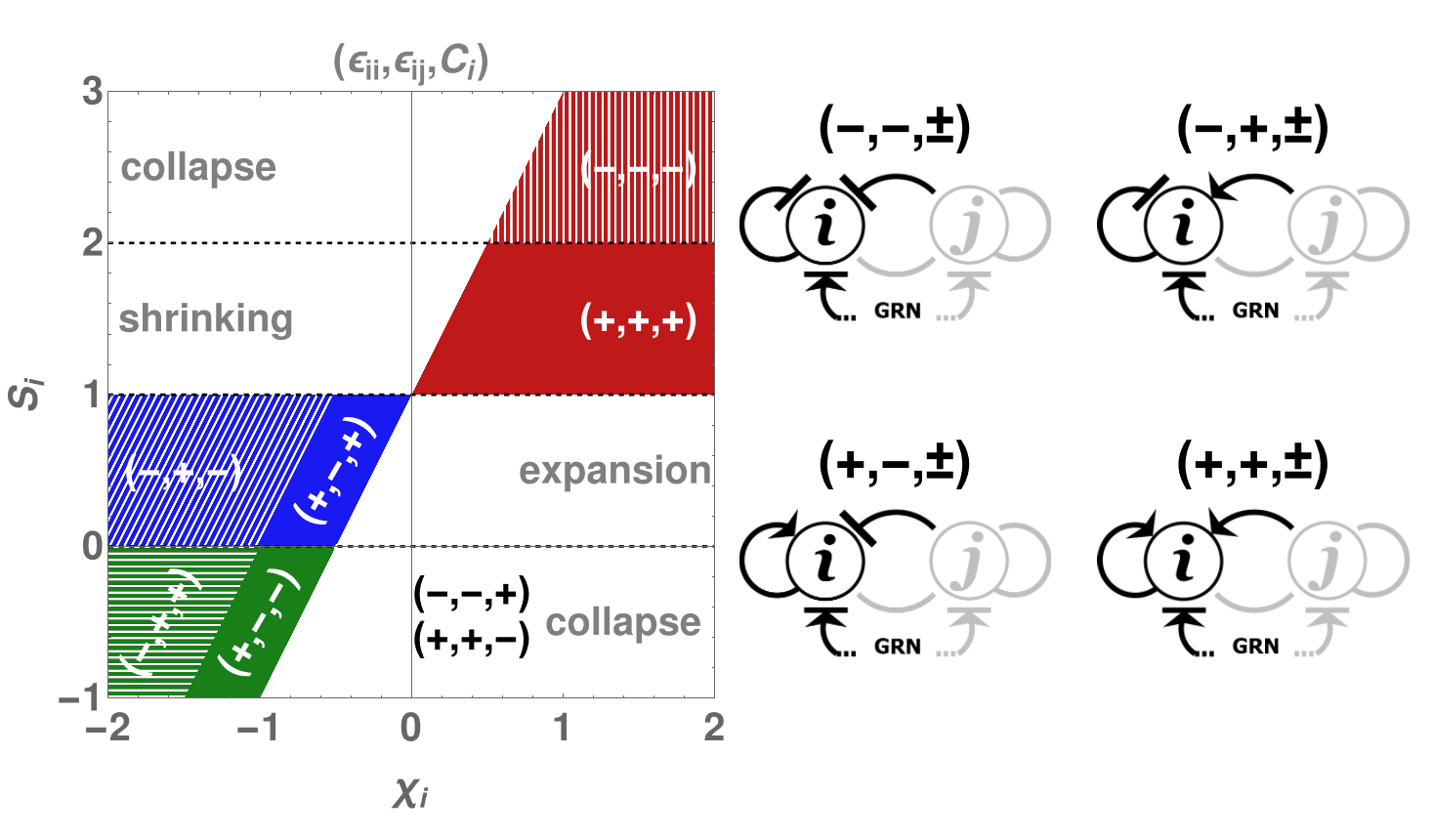}
\caption{Half-motifs $(\epsilon_{ii},\epsilon_{i\neq j},C_i)$ and their stability regions. Colored fields show stability regions for half-motifs with auto-activation ($\epsilon_{ii}>0$), which overlap with stability regions for auto-repressive half-motifs ($\epsilon_{ii}<0$, dashed fields). \label{fig:halfmotif} }
\end{figure}
%============================

These stability conditions \eqref{eq:stab_cond} are valid in the entire range of $S_i$ and $\chi_i$, not only for $0<S_i<1$ and $\chi_i<0$. This allows us to discuss the stability of all 64 two-gene network motifs, described by six constants (four $\epsilon_{ij}$ and two $C_i$), which can be either $>0$ or $<0$. We will call a trio $(\epsilon_{ii}, \epsilon_{i\neq j},C_i)$ together with node $i$ the `half-motif', as two such trios form a full two-gene motif. There are only 8 half-motifs. $(\epsilon_{ii}, \epsilon_{i\neq j}, C_i)$ can be mapped into the pair of $(\chi_i,S_i)$ via Eqs. \eqref{eq:params}. In stable systems, parameters must be chosen in such a way that point $(\chi_1,S_1)$ corresponds to point $(\chi_2,S_2)$ with the same $\Delta X$. In Fig. \ref{fig:phase_trans}D we illustrate the SGEP phase diagram of $\Delta X(\chi_i,S_i)$. The phase diagram is symmetric about the point $(0,1)$, so, having $(\chi_1,S_1)$ fixed, $(\chi_2,S_2)$ can be chosen in two ways. In order to classify the stability of resulting pattern it is enough to check to which part of phase diagram in Fig.~\ref{fig:phase_trans}D each of the half-motifs constituting the full network motif can be assigned. SGEPs are formed only from the two potentially stable half-motifs.

In Fig.~\ref{fig:halfmotif} the stability regions of all possible half-motifs are shown. For half-motifs with $C_i>0$, stabilization is the result of competition between interactions and the spontaneous behavior of domains. For $(+,+,+)$ and $(-,+,+)$ activation is counterbalanced by the spontaneous shrinking or collapsing of domains ($S_i>1$ and $S_i<0$, respectively). For these cases, the boundaries of initial domains must be placed within $\lambda_i$'s range, so the stabilization precedes their spontaneous decay. The half-motif $(+,-,+)$ is less restrictive for initial conditions, as it involves a spontaneously growing domain ($0<S_i<1$). Combinations of $(+,-,+)$ with $(-,+,+)$ are often encountered in biological systems, such as spinal cord development \cite{bib:Balaskas_2012}, limb formation \cite{bib:Raspopovic_2014,bib:Marcon_2016,bib:Landge_2020,bib:Mateus_2020} and segmentation in the \textit{Drosophila} embryo \cite{bib:Verd_2017,bib:Verd_2019}.

For half-motifs with $C_i<0$ stabilization is the result of competition between the default activation ($C_i<0$) and inhibiting interactions. There are no restrictions for initial conditions, as genes with $C_i<0$ are spontaneously expressed in undifferentiated tissue. 

Two half-motifs cannot be stabilized: $(-,-,+)$ and $(+,+,-)$. The former has no activating interactions, thus it cannot sustain expression in the long run. Conversely, the latter is activated by default and by both interactions, thus it spreads in unbounded manner. Interestingly, half-motif $(-,-,C_i)$ is found in many biological systems \cite{bib:Alon_2007,bib:Balaskas_2012,bib:Verd_2017,bib:Verd_2019}, but our results show that it requires external activation ($C_i<0$) to ensure stability. 

Finally, the stability of half-motifs with $\epsilon_{ii}<0$ has additional limitation. In the two-gene system, the effective threshold for activation reads $\tilde C_i=C_i-\epsilon_{ij}\psi_j$. Similar to one-component systems, when $\tilde C_i<0$ and $ \epsilon_{ii}\tilde \psi_i<\tilde C_i$, the production of $i$-th gene tends to $\tilde C_i/\epsilon_{ii}$ instead of $\tilde \psi_i$. This causes the stability conditions to fail when $S_i<2+2\chi_i$ for $(-,+,\pm)$ and when $S_i<2$ for $(-,-,-)$.

We have shown that the quasi-non-linear model provides an in-depth insight into the problem of GEP stabilization. It elucidates the single-gene dynamics and demonstrates that the formation of biologically relevant GEPs (traveling or genuinely stable) is governed by a phase transition. Further, it provides stability conditions for two-gene network motifs encountered in developmental GRNs. Our predictions should hold for systems with additional genes and interactions, as long as contact zones between domains are separated by at least $\lambda_i$, which allows us to consider them separately. The model also exhibits much potential for further generalizations. 

\begin{acknowledgments}
We thank T. Sokolowski and B. Waclaw for comments on the manuscript. M.M. and M.Z. were supported by the Polish National Agency for Academic Exchange. R.H. and M.Z. were supported by the Narodowe Centrum Nauki, Poland (SONATA, 2017/26/D/NZ2/00454).
\end{acknowledgments}

\end{document}

\maketitle

\tableofcontents

\pagebreak
\section{Parameters for figures}
All figures and simulations were performed with Wolfram Mathematica 12.1.

\subsection{Figure 1}

Figures (a)-(e) were generated using the following common parameters:

\begin{center}
\begin{tabular}{|c|c|c|c|c|c|c|c|c|c|c|c|c|c|}
\hline
$D_1$  & $D_2$  & $\gamma_1$ & $\gamma_2$  &  $H_1$  & $H_2$  & $\epsilon_{11}$ & $\epsilon_{22}$ & $X_1(0)$ & $X_2(0)$ & $\lambda_1$ & $\lambda_2$ & $\sigma_1$ & $\sigma_2$\\
\hline
\multicolumn{2}{|c|}{[$\mu m/s^2$]} & \multicolumn{4}{c|}{[$s^{-1}$]} & \multicolumn{2}{c|}{-} & \multicolumn{4}{c|}{[$\mu m$] } & \multicolumn{2}{c|}{-}\\
\hline
1.0 & 1.0 & 0.0005 & 0.0004 & 0.01 & 0.02 & 1.0 & 1.0 & -187.5 & 187.5 & 50 & 44.72 & -1 & +1\\
\hline
\end{tabular}
\end{center}
where $\gamma_i$ and $D_i$ were chosen to fit into a typical range of biologically relevant parameters (see e.g. A. Kicheva et al., Curr. Opin. Genet. Dev. 22, 6, 527 (2012)). The following table shows the parameters $C_i$ and $\epsilon_{i\neq j}$ used for each plot as well as the values of $S_i$, $R_i$ and $\Delta X$ corresponding to these parameters:

\begin{center}
\begin{tabular}{|c|c|c|c|c|c|c|c|c|c|c|}
\hline
Panel & $C_1$ & $C_2$ & $\epsilon_{12}$ & $\epsilon_{21}$ & $S_1$ & $S_2$ & $R_1$ & $R_2$ & $\Delta X$ [$\mu m$] \\
\hline
(a) & 3.5 & 3.5 & -0.21 & -0.37 & 0.35 & 0.28 & 1.476 & 1.432 & - \\
(b) & 3.5 & 6.0 & -0.2 & -1.0 & 0.35 & 0.24 & 0.3 & 0.9 & -\\
(c) & 4.0 & 6.5 & -0.45 & -2.1 & 0.4 & 0.26 & -0.467 & -0.119 & - \\
(d) & 4.5 & 7.5 & -0.147& -1.137& 0.45 & 0.3 & 0.5 & 0.539 & -34.657 \\
(e) & 4.5 & 7.5 & 0.641& -4.875& 0.45 & 0.3 & -0.6 & -0.55 & 45.814\\
\hline
\end{tabular}
\end{center}
For panels (d) and (e) the value of $R_1$ was treated as input, while $\epsilon_{12}$, $R_2$ and $\epsilon_{21}$ were calculated to satisfy stabilization conditions. 

\subsection{Figure 3}
\subsubsection{Panel (a)}
The heatmap representation of the order parameter was generated for the two-gene system with the following parameters:
\begin{center}
\begin{tabular}{|c|c|c|c|c|c|c|c|c|c|c|c|c|c|}
\hline
$D_1$  & $D_2$  & $\gamma_1$ & $\gamma_2$  &  $H_1$  & $H_2$  & $\epsilon_{11}$ & $\epsilon_{22}$ & $X_1(0)$ & $X_2(0)$ & $\lambda_1$ & $\lambda_2$ & $\sigma_1$ & $\sigma_2$\\
\hline
\multicolumn{2}{|c|}{[$\mu m/s^2$]} & \multicolumn{4}{c|}{[$s^{-1}$]} & \multicolumn{2}{c|}{-} & \multicolumn{4}{c|}{[$\mu m$] } &\multicolumn{2}{c|}{- }\\
\hline
1.0 & 1.0 & 0.0005 & 0.0004 & 0.01 & 0.02 & 1.0 & 1.0 & -75 & 75 & 50 & 44.72 & -1 & +1\\
\hline
\end{tabular}
\end{center}
Interaction constants  $\epsilon_{12}$, $\epsilon_{21}$ were chosen to cover the phase space presented in the plot. System size was 5000[$\mu m$] and the simulation lasted for 36 000[s]. Front positions $X_i(t)$ were found numerically, by solving the eq. (7) from the main text. $X_i(t)$'s corresponding to the final 25\% of simulation period were fitted with a linear model to obtain asymptotic velocity $v_i$. We checked that within the simulation time fronts did not get within the range of interaction with the system boundary and the constant velocity dynamics was well-established in the fitted range of data. 

\begin{figure}[h]
\centering
\includegraphics[width=0.5\textwidth]{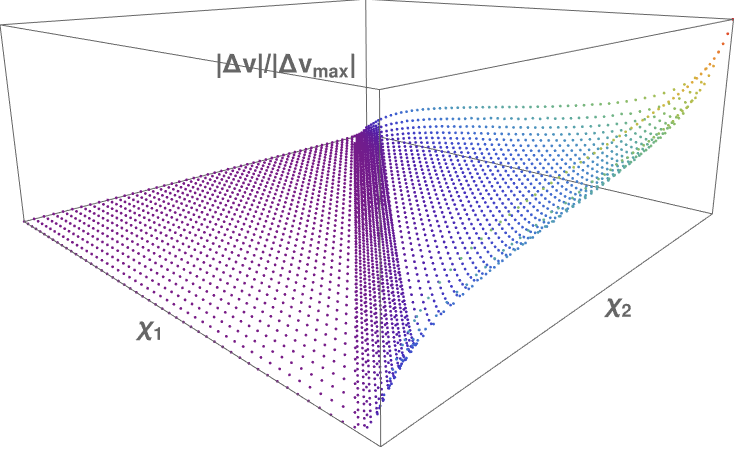}
\caption{3D presentation of order parameter over $(\chi_1,\chi_2)$ plane (compare Fig. 3(a) from the main text). \label{fig:3d_order}}
\end{figure}

\subsubsection{Panels (b) and (c):}
The following parameters were used to generate these plots:

\begin{center}
\begin{tabular}{|c|c|c|c|c|c|}
\hline
$D_1$ [$\mu m/s^2$] & $D_2$ [$\mu m/s^2$]  & $\gamma_1$ [$s^{-1}$]& $\gamma_2$ [$s^{-1}$] &  $S_1$ & $S_2$\\
\hline
1.0 & 1.0 & 0.0005 & 0.0004 & 0.05 & 0.1 \\
\hline
\end{tabular}
\end{center}

\section{Estimation of velocity of domain shift in \textit{Drosophila}}

In order to estimate the biologically relevant part of the low-velocity plateau in TGEP phase, we estimate the velocity of domain boundaries from reported shifts in gap gene expression domains in \textit{Drosophila melanogaster} (see B. Verd et al., PLoS Comput. Biol. 13, (2017)). We consider boundary between Knirps (Kni) and Giant (Gt), that is located in 73.5\% of the relative A-P position at developmental time class C14A-T1, and in 77\% of the relative A-P position at C14A-T8 (see B. Verd et al., PLoS Comput. Biol. 13, (2017)). Hence the Kni-Gt boundary shifts in the anterior direction by 3.5\% of the relative A-P position from time class T1 to T8. The AP size of the embryo in absolute units is close to 500 $\mu$m (see T. A. Markow et al., J. Evol. Biol. 22, 430 (2009)) and time elapsing from T1 to T8 is about 44 minutes (see J. Jaeger et al., Genetics 167, 1721 (2004)). Thus, the estimated velocity of Kni-Gt is 17.5 $\mu$m over 44 minutes, that corresponds to $v = 24$ $\mu$m/h. In Fig. 3(b) of the main text the region with $v=\pm$24 $\mu$m/h corresponds to the nearest color-coded regions adjacent to the SGEP line ($v = 0$).

%========DERIVATION=================
\section{Full derivation of the stabilization conditions}
\subsection{Preliminary steps}
We consider the following model:
 \begin{equation}
\left\{ \begin{gathered}
\partial_t \psi_1(x,t)=D_1\partial_{xx}\psi_1(x,t)-\gamma_1 \psi_1(x,t)+H_1\theta(F_1(\psi_1,\psi_2))\\
\partial_t \psi_2(x,t)=D_2\partial_{xx} \psi_2(x,t)-\gamma_2 \psi_2(x,t)+H_2\theta(F_2(\psi_1,\psi_2))
\end{gathered}\right. \label{eq:main}
 \end{equation}
with initial conditions given by:
 \begin{equation}
\psi_i(x,0)=A_i \theta\left(\sigma_i(x-X_i(0))\right)
 \end{equation}
$\sigma_i$ indicates the orientation of the domain. $\sigma_i=-1$ corresponds to the domain occupying the region $x\in(-\infty,X_i(t)]$ and $\sigma_i=+1$ to $x\in [X_i(t),+\infty)$. 

Up to the knowledge of position of the activation front, $X_i(t)$, the solution of  system \eqref{eq:main} can be obtained with standard methods for linear inhomogeneous partial differential equations:
  \begin{equation}
\left\{ \begin{gathered}
\psi_1(x,t)=\int_{-\infty}^{+\infty}dx'G_1(x-x',t)\psi_1(x',0)+H_1\sigma_1\int_0^tdt'\int^{\sigma_1\infty}_{X_1(t')} dx' G_1(x-x',t-t') \\
\psi_2(x,t)=\int_{-\infty}^{\infty}dx'G_2(x-x',t)\psi_2(x',0)+H_2\sigma_2\int_0^tdt'\int_{X_2(t')}^{\sigma_2\infty} dx' G_2(x-x',t-t') 
\end{gathered}\right. \label{eq:solution}
 \end{equation}
 where:
 \begin{equation}
 G_i(\Delta x, \Delta t)=\frac{e^{-\gamma_i \Delta t-\frac{\Delta x^2}{4D_i \Delta t}}}{\sqrt{4\pi D_i \Delta t}}
 \end{equation}
 
The positions of the activation fronts $X_i(t)$ are defined by the equations:
 \begin{equation}
 \begin{gathered}
C_1= \epsilon_{11} \psi_1(X_1(t),t)+ \epsilon_{12} \psi_2(X_1(t),t)\\
C_2= \epsilon_{21}\psi_1(X_2(t),t)+ \epsilon_{22} \psi_2(X_2(t),t)
 \end{gathered}
 \end{equation}
For the time scale $t\gg \max(\gamma_1^{-1},\gamma_2^{-1})$ we can neglect the influence of initial conditions as these terms diminish as $e^{-\gamma_i t}$. Then, the equations simplify into:
  \begin{equation}
 \begin{gathered}
C_1=\epsilon_{11} H_1\sigma_1\int_0^tdt'\int^{\sigma_1\infty}_{X_1(t')} dx' G_1(X_1(t)-x',t-t') + \epsilon_{12} H_2\sigma_2\int_0^tdt'\int_{X_2(t')}^{\sigma_2\infty} dx' G_2(X_1(t)-x',t-t')\\
C_2=\epsilon_{21} H_1\sigma_1\int_0^tdt'\int^{\sigma_1\infty}_{X_1(t')} dx' G_1(X_2(t)-x',t-t') + \epsilon_{22} H_2\sigma_2\int_0^tdt'\int_{X_2(t')}^{\sigma_2\infty} dx' G_2(X_2(t)-x',t-t')
 \end{gathered}\label{eq:integral_form}
 \end{equation}
The integrals over $x'$ can be performed outright to obtain:
 \begin{equation}
 \begin{gathered}
 \begin{split}
C_1=& \frac{\epsilon_{11} H_1}{2}\int_0^tdt' e^{-\gamma_1(t-t')} \left(1-\sigma_1\textrm{Erf}\left( \frac{X_1(t')-X_1(t)}{\sqrt{4 D_1 (t-t')}}\right) \right)+\\
&+\frac{\epsilon_{12}H_2}{2}\int_0^tdt' e^{-\gamma_2(t-t')}\left( 1-\sigma_2\textrm{Erf}\left( \frac{X_2(t')-X_1(t)}{\sqrt{4 D_2 (t-t')}}\right) \right)
 \end{split}\\
 \begin{split}
C_2= & \frac{\epsilon_{21} H_1}{2}\int_0^tdt' e^{-\gamma_1(t-t')} \left(1-\sigma_1\textrm{Erf}\left( \frac{X_1(t')-X_2(t)}{\sqrt{4 D_1 (t-t')}}\right) \right)+\\
&+\frac{\epsilon_{22}H_2}{2}\int_0^tdt' e^{-\gamma_2(t-t')}\left( 1-\sigma_2\textrm{Erf}\left( \frac{X_2(t')-X_2(t)}{\sqrt{4 D_2 (t-t')}}\right) \right)
 \end{split}
 \end{gathered}
 \end{equation}
Further, we can perform the simple integrals over exponential terms:
\begin{equation}
\int_0^{+\infty} dt' e^{-\gamma_i(t-t')}=\gamma_i^{-1}(1-e^{-\gamma_it})\xrightarrow{t\gg\max(\gamma_1^{-1},\gamma_2^{-1})}\gamma_i^{-1}
\end{equation}
 Thus, for $t\gg \max(\gamma_1^{-1},\gamma_2^{-1})$, the equations take the form:
 \begin{equation}
 \begin{gathered}
 \begin{split}
C_1=& \frac{\epsilon_{11} H_1}{2\gamma_1}-\frac{\epsilon_{11} H_1}{2}\sigma_1\int_0^tdt' e^{-\gamma_1(t-t')} \textrm{Erf}\left( \frac{X_1(t')-X_1(t)}{\sqrt{4 D_1 (t-t')}}\right)\\
&+\frac{\epsilon_{12}H_2}{2\gamma_2}-\frac{\epsilon_{12}H_2}{2}\sigma_2\int_0^tdt' e^{-\gamma_2(t-t')}\textrm{Erf}\left( \frac{X_2(t')-X_1(t)}{\sqrt{4 D_2 (t-t')}}\right)
 \end{split}\\
 \begin{split}
C_2= &\frac{\epsilon_{21} H_1}{2\gamma_1}-\frac{\epsilon_{21} H_1}{2}\sigma_1\int_0^tdt' e^{-\gamma_1(t-t')} \textrm{Erf}\left( \frac{X_1(t')-X_2(t)}{\sqrt{4 D_1 (t-t')}}\right)+\\
&+\frac{\epsilon_{22}H_2}{2\gamma_2}-\frac{\epsilon_{22}H_2}{2}\sigma_2\int_0^tdt' e^{-\gamma_2(t-t')}\textrm{Erf}\left( \frac{X_2(t')-X_2(t)}{\sqrt{4 D_2 (t-t')}}\right)
 \end{split}
 \end{gathered}
 \end{equation}
 We can now introduce the parameters:
 \begin{align}
 S_i=\frac{2C_i\gamma_i}{\epsilon_{ii}H_i}&&\chi_i=\frac{\epsilon_{ij}H_j\gamma_i}{\epsilon_{ii}H_i\gamma_j}
\end{align}
 and, after dividing both sides by $\frac{2\gamma_i}{\epsilon_{ii}H_i}$, rewrite the system as:
 \begin{equation}
 \begin{gathered}
 \begin{split}
&S_1-1-\chi_1=\\
& -\sigma_1\gamma_1\int_0^tdt' e^{-\gamma_1(t-t')} \textrm{Erf}\left( \frac{X_1(t')-X_1(t)}{\sqrt{4 D_1 (t-t')}}\right)-\chi_1\sigma_2\gamma_2\int_0^tdt' e^{-\gamma_2(t-t')}\textrm{Erf}\left( \frac{X_2(t')-X_1(t)}{\sqrt{4 D_2 (t-t')}}\right)
 \end{split}\\
 \begin{split}
&S_2-1-\chi_2=\\ 
&-\chi_2\sigma_1\gamma_1\int_0^tdt' e^{-\gamma_1(t-t')} \textrm{Erf}\left( \frac{X_1(t')-X_2(t)}{\sqrt{4 D_1 (t-t')}}\right)-\sigma_2\gamma_2\int_0^tdt' e^{-\gamma_2(t-t')}\textrm{Erf}\left( \frac{X_2(t')-X_2(t)}{\sqrt{4 D_2 (t-t')}}\right)
 \end{split}
 \end{gathered} \label{eq:ready_form}
 \end{equation}
 
 \subsection{Solving the one-component system}
\begin{figure}
\centering
\includegraphics[width=0.5\textwidth]{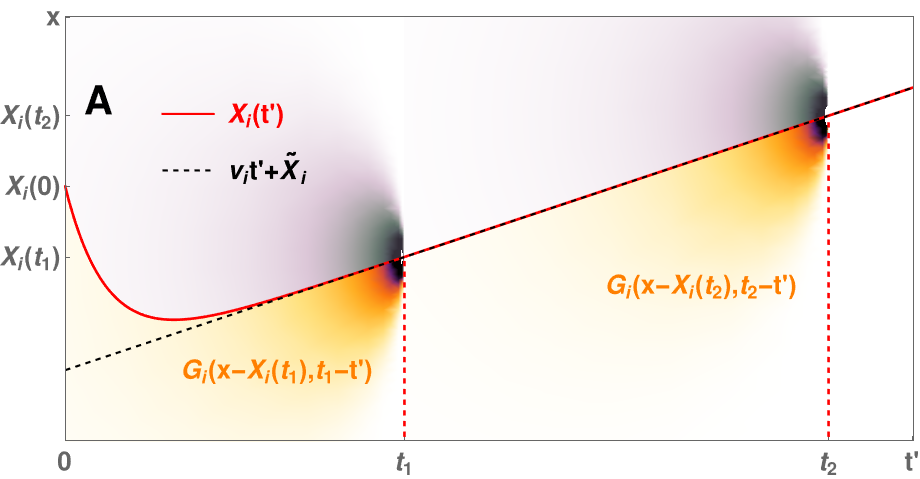}
\caption{The origin of the constant velocity ansatz in one-component case. The density map in the background is the function $G_i(x-X_i(t),t-t')$ show for two moments, $t'=t_1$ and $t'=t_2$. The color-shaded region is the area of integration in the equation \eqref{eq:one_comp}. When the interaction region is translated alongside the straight line $x=v_it'+const.$ (black, dashed line), the value of of the integral is conserved. This figure is the same as Fig. 2(a) in the main text. \label{fig:int1}}
\end{figure}

Let us get back to the equations in the form \eqref{eq:integral_form} and assume no cross-interactions, $\epsilon_{i\neq j}=0$.  Thus, we can consider each component separately. The dynamics of the $i$-th front is governed by the equation:
\begin{equation}
\frac{C_i}{\epsilon_{ii}H_i}=\sigma_i\int_0^t dt'\int^{\sigma_i\infty}_{X_i(t')}dx' G_i(x'-X_i(t),t-t') \label{eq:one_comp}
\end{equation}
The physical meaning of this equation is that $X_i(t)$ must evolve in such way that the space-time integral of function $G_i$ on the right-hand side of \eqref{eq:one_comp} is conserved and equal to the constant on the left-hand side of \eqref{eq:one_comp}. This integral is taken over the space-time area between the line of current moment $t'=t$ and the past positions of the front, i.e. $X_i(t'<t)$. This is illustrated in Fig. \ref{fig:int1}. Most of the integral value is concentrated in the vicinity of $x'= X_i(t)$, as this is where $G_i$ is centered.

%=======fig:int1================

One can observe now that the value of the integral is conserved when translated along a straight line in the time-space, i.e. the front moves with the constant velocity:
\begin{equation}
X_i(t)=v_it+\tilde X_i
\end{equation}

We can now insert this result into \eqref{eq:ready_form} (taken for $\chi_i=0$, no cross-interactions) and change the variables to $\Delta t=t-t'$ This results in the equation:
\begin{equation}
S_i-1=\sigma_i\gamma_i\int_0^t dt' e^{-\gamma_i\Delta t}\textrm{Erf}\left(\frac{v_i\sqrt{\Delta t}}{\sqrt{4D_i}} \right)=\sigma_i \gamma_i I_1\left(\gamma_i,\frac{v_i}{\sqrt{4D_i}},t \right)
\end{equation}
The integral of the type $I_1(\gamma_i,b,t)$ is calculated in the last section (formula \eqref{eq:I1}). Employing this result and taking the $t \to +\infty $ limit, we obtain:
\begin{equation}
S_i-1=\sigma_i \frac{v_i}{\sqrt{4D_i\gamma_i+v_i^2}} \label{eq:v_raw}
\end{equation}
which is formula (10) from the main text. We can use equation \eqref{eq:v_raw} to decode the sign of $v_i$ for each front:
\begin{equation}
\begin{array}{c|cc}
&\sigma_i=-1&\sigma_i=1\\
\hline
S_i>1&v_i<0&v_i>0\\
S_i<1&v_i>0&v_i<0\\
\end{array} \label{eq:1dtab}
\end{equation}
This shows that, indeed, in the regime of front dynamics ($0<S_i<2$), $S_i$ controls whether the domain expands ($S_i<1$) or shrinks ($S_i>1$). Eq. \eqref{eq:v_raw} can be easily solved for $v_i^2$, leading to:
\begin{equation}
v_i^2=4D_i\gamma_i\frac{(S_i-1)^2}{1-(S_i-1)^2}
\end{equation}
From this expression and using \eqref{eq:1dtab}, it is possible to obtain $v_i$, which is shown in Fig. 2(b), in the main text.

\subsection{Solving the two-component system in TGEP phase}
\begin{figure}
\centering
\includegraphics[width=0.95\textwidth]{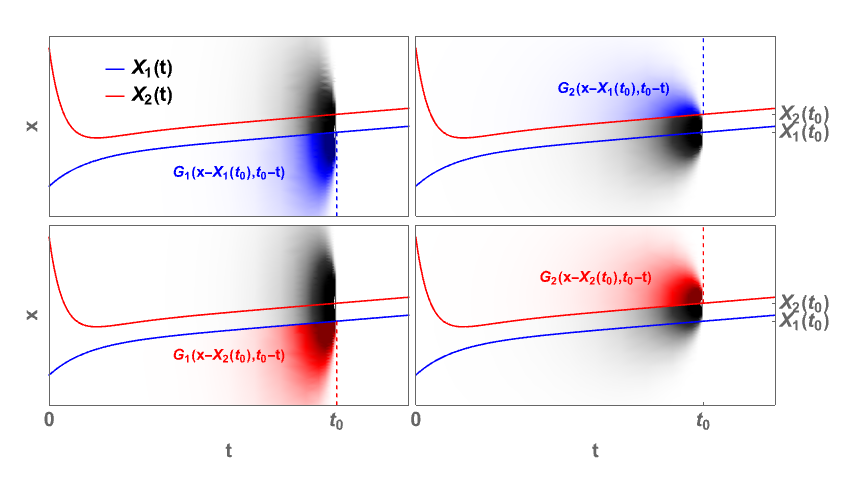}
\caption{The origin of constant velocity ansatz in two-component case. The density maps in the background show the functions $G_i(x-X_i(t),t-t')$, the color-shaded regions indicate the actual area of integration in \eqref{eq:integral_form}. Diagonal panels represent the auto-interaction integrals and off-diagonal panels show cross-interaction integrals. Upper row (blue shading) corresponds to $i=1$ and lower (red shading) to $i=2$. One can notice that translating the entire system along a straight line $x=v t'$ asymptotically preserves the values of all integrals. Thus, for interacting system both fronts must move with constant and the same velocity. \label{fig:int2}}
\end{figure}

Let us now focus now on the interacting variant of \eqref{eq:integral_form}, i.e. $\epsilon_{i\neq j}\neq 0$. In this case, the evolution of $X_1(t)$ and $X_2(t)$ must be such that the sum of two integral terms (one for the auto-interaction and the other for the cross-interaction) must preserve the constant value. Inspecting the graphical representation of \eqref{eq:integral_form} (Fig. \ref{fig:int2}) one can notice that this is achieved for two fronts traveling with \textit{the same} velocity, i.e $v_1=v_2=v$. For $v_1\neq v_2$ the cross-interaction integrals (off-diagonal panels in Fig.~\ref{fig:int2}) would have to change their values in time. Thus, the stable solution for two fronts remaining within their interaction range requires them to adopt the same velocity. 

%========fig 2 comp ============
%============================

Getting back to \eqref{eq:ready_form}, we can insert the constant velocity ansatz:
\begin{align}
X_i(t)=vt+\tilde X_i&&\Delta X=\tilde X_2-\tilde X_1
\end{align}
into these equations. Changing the variables to $\Delta t=t-t'$, we obtain:
\begin{equation}
 \begin{gathered}
 \begin{split}
&S_1-1-\chi_1=\\
&\sigma_1\gamma_1\int_0^{t }d\Delta t e^{-\gamma_1\Delta t} \textrm{Erf}\left( v\sqrt{\frac{\Delta t}{4 D_1 }}\right)+\chi_1\sigma_2\gamma_2\int_0^{t}d\Delta t e^{-\gamma_2\Delta t}\textrm{Erf}\left( v\sqrt{\frac{\Delta t}{4 D_2 }}-\frac{\Delta X}{\sqrt{4D_2\Delta t}}\right)
 \end{split}\\
 \begin{split}
 &S_2-1-\chi_2=\\
 & \chi_2 \sigma_1\gamma_1 \int_0^{t} d\Delta t e^{-\gamma_1 \Delta t} \textrm{Erf}\left( v\sqrt{\frac{\Delta t}{4 D_1}}+\frac{\Delta X}{\sqrt{4D_1\Delta t}}\right)+\sigma_2\gamma_2\int_0^{t}d\Delta t e^{-\gamma_2\Delta t}\textrm{Erf}\left( v\sqrt{\frac{\Delta t}{4 D_2}}\right) 
\end{split}
 \end{gathered}
 \end{equation}
We can rewrite this system using the integrals $I_1(\gamma_i,b,t)$ (formula \eqref{eq:I1}) and $I_3(\gamma_i,a,b,t)$ (formula \eqref{eq:I3}):
\begin{equation}
 \begin{gathered}
 \begin{split}
&S_1-1-\chi_1=\sigma_1\gamma_1 I_1\left(\gamma_1,\frac{v}{\sqrt{4D_1}},t\right)+\chi_1\sigma_2\gamma_2I_3\left(\gamma_2,-\frac{\Delta X}{\sqrt{4D_2}},\frac{v}{\sqrt{4D_2}},t\right)
 \end{split}\\
 \begin{split}
 &S_2-1-\chi_2=\sigma_1\chi_2\gamma_1I_3\left(\gamma_1,\frac{\Delta X}{\sqrt{4D_1}},\frac{v}{\sqrt{4D_1}},t\right)+\sigma_2\gamma_2 I_1\left(\gamma_2,\frac{v}{\sqrt{4D_2}}\right)
\end{split}
 \end{gathered}
 \end{equation}
We can now now employ the explicit forms of integrals $I_1$ and $I_3$. To obtain the long term solution, we must calculate the limits:
\begin{gather}
\begin{split}
\lim_{t\to+\infty}I_1\left(\gamma_i,\frac{v}{\sqrt{4D_i}},t \right)=\frac{v}{\gamma_i\sqrt{4D_i}}\frac{1}{\sqrt{\frac{v^2}{4D_i}+\gamma_i}}=\frac{v}{\gamma_i\sqrt{v^2+4D_i\gamma_i}}
\end{split}\\
\begin{split}
&\lim_{t\to+\infty}I_3\left(\gamma_i,\frac{\pm\Delta X}{\sqrt{4D_i}},\frac{v}{\sqrt{4D_i}},t \right)=\\
&=\frac{\textrm{sgn}(\pm\Delta X)}{\gamma_i}-0+\frac{e^{\mp\frac{v\Delta X}{2D_i}-\frac{|\Delta X|\sqrt{4D_i\gamma_i+v^2}}{2D_i}}}{\gamma_i}\left(\frac{v}{\sqrt{4D_i \gamma_i +v^2}}-\textrm{sgn}(\pm\Delta X) \right)-0
\end{split}
\end{gather}
Eventually, we obtain the system of equations:
\begin{equation}
 \begin{gathered}
 \begin{split}
&S_1-1-\chi_1=\\
&\sigma_1\frac{v}{\sqrt{4D_1 \gamma_1+v^2}}+\chi_1\sigma_2\left[-\textrm{sgn}(\Delta X)+e^{\frac{v\Delta X}{2D_2}-\frac{|\Delta X|\sqrt{4D_2\gamma_2+v^2}}{2D_2}}\left(\frac{v}{\sqrt{4D_2 \gamma_2 +v^2}}+\textrm{sgn}(\Delta X) \right)\right] \end{split}
\\
\begin{split}
&S_2-1-\chi_2=\\
&\chi_2\sigma_1\left[\textrm{sgn}(\Delta X)+e^{-\frac{v\Delta X}{2D_1}-\frac{|\Delta X|\sqrt{4D_1\gamma_1+v^2}}{2D_1}}\left(\frac{v}{\sqrt{4D_1 \gamma_1 +v^2}}-\textrm{sgn}(\Delta X) \right)\right]+\sigma_2\frac{v}{\sqrt{v^2+4D_2 \gamma_2}} 
\end{split} 
\end{gathered} \label{eq:main_fin}
 \end{equation}
which is equivalent to the formula (11) from the main text.

The stability conditions are obtained by assuming that $v=0$. Then the system can be transformed into the following form:
\begin{equation}
\begin{gathered}
S_1-1-\chi_1=\chi_1\sigma_2\left(-\textrm{sgn}(\Delta X)+ \textrm{sgn}(\Delta X) e^{-|\Delta X|/\lambda_2}\right)\\
S_2-1-\chi_2=\chi_2\sigma_1\left(\textrm{sgn}(\Delta X)-\textrm{sgn}(\Delta X) e^{-|\Delta X|/\lambda_1}\right)
\end{gathered}\label{eq:main_stable}
\end{equation}
We can disentangle $|\Delta X|$ from each equation:
\begin{equation}
\begin{gathered}
|\Delta X|=-\lambda_2\ln \left(1+\frac{1}{\sigma_2\textrm{sgn}(\Delta X)}\left(\frac{S_1-1}{\chi_1}-1\right) \right)\\
|\Delta X|=-\lambda_1\ln \left(1-\frac{1}{\sigma_1\textrm{sgn}(\Delta X)}\left(\frac{S_2-1}{\chi_2}-1\right) \right)
\end{gathered}
\end{equation}
The system has a solution provided that the $\Delta X$ defined by one equation is equal to the $\Delta X$ defined by the other equation. Thus, it must be satisfied that:
\begin{equation}
\left(1+\frac{1}{\sigma_2\textrm{sgn}(\Delta X)}\left(\frac{S_1-1}{\chi_1}-1\right) \right)^{-\lambda_2}=\left(1-\frac{1}{\sigma_1\textrm{sgn}(\Delta X)}\left(\frac{S_2-1}{\chi_2}-1\right) \right)^{-\lambda_1} \label{eq:stab_raw}
\end{equation}
Which is the raw form of stabilization conditions. This formula can be further reduced when we introduce the effective variable $R_i$:
\begin{equation}
R_i=\frac{S_i-1}{\chi_i}-1
\end{equation}
Then:
\begin{equation}
\begin{gathered}
|\Delta X|=-\lambda_2 \ln \left(1+\frac{1}{\sigma_2\textrm{sgn}(\Delta X)} R_1\right) \\
|\Delta X|=-\lambda_1 \ln \left(1-\frac{1}{\sigma_1\textrm{sgn}(\Delta X)} R_2 \right)
\end{gathered}\label{eq:DeltaX1}
\end{equation}
The right-hand side of these equations is non-negative, thus these equations can be satisfied only for:
\begin{align}
\frac{1}{\sigma_2 \textrm{sgn}(\Delta X)}R_1\in[-1,0]&&\land&&-\frac{1}{\sigma_1 \textrm{sgn}(\Delta X)}R_2\in[-1,0]
\end{align}
 This means that:
\begin{align}
R_i\in[-1,1]&&\textrm{sgn}(R_1)=-\sigma_2\textrm{sgn}(\Delta X)&&\textrm{sgn}(R_2)=\sigma_1\textrm{sgn}(\Delta X) \label{eq:existence}
\end{align}
This can be abbreviated to:
\begin{align}
\sigma_i\textrm{sgn}(R_i)=-\sigma_j\textrm{sgn}(R_j)&&\textrm{sgn}(\Delta X)=(-1)^i\sigma_j \textrm{sgn}(R_i)
\end{align}
Rewriting $R_i=\textrm{sgn}(R_i)|R_i|$ and using relations \eqref{eq:existence}, formulas \eqref{eq:DeltaX1} can be expressed as:
\begin{equation}
|\Delta X|=-\lambda_j \ln \left(1-|R_i| \right) 
\end{equation}
and \eqref{eq:stab_raw} as:
\begin{equation}
(1-|R_1|)^{-\lambda_2}=(1-|R_2|)^{-\lambda_1}
\end{equation}
These are the stability conditions as provided in the main text.

\subsection{Solving the two-component system in IGEP phase}
\begin{figure}
    \centering
    \includegraphics[width=0.9\textwidth]{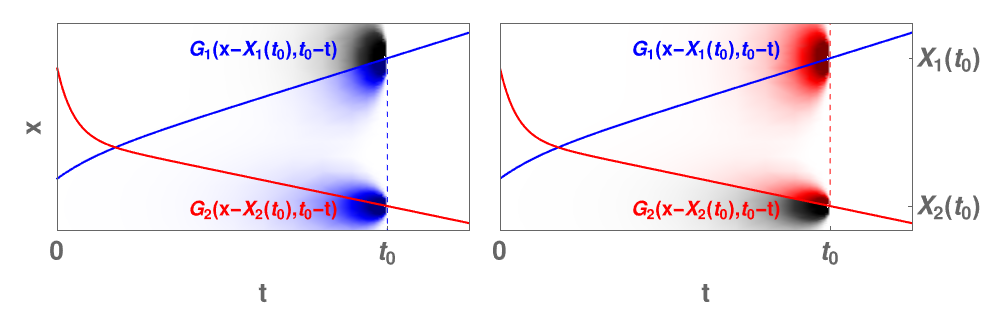}
    \caption{Illustration of auto- and cross-interaction integrals from Eqs. \eqref{eq:ready_form} in the IGEP phase. When domains overlap, but fronts are beyond interaction range, the cross-interaction integrals saturate at constant value.
    \label{fig:IGEP_ints}}
\end{figure}

In the IGEP phase we assume that two domains began to inter-grow, increasing the overlap. When activation fronts are far beyond interaction range ($\gg \lambda_i$), the cross-interaction integrals in \eqref{eq:ready_form} saturate at certain constant values. This is illustrated in Fig. \ref{fig:IGEP_ints}. In this state, the dynamics resembles the case of non-interacting genes, but the equations of motion will be modified by the additional constant term, coming from the cross-interaction with the `background' gene.

Once again we introduce the constant velocity ansatz, in the form:
\begin{align}
X_i(t)=v_i t+\tilde X_i&&\Delta X=\tilde X_2-\tilde X_1&&\Delta v=v_2-v_1
\end{align}
where $v_1\neq v_2$. Inserting this ansatz in \eqref{eq:ready_form} allows us to rewrite eqs. \eqref{eq:ready_form} in the following manner:
 \begin{equation}
 \begin{gathered}
 \begin{split}
&S_1-1-\chi_1=\\
& \sigma_1\gamma_1\int_0^tdt' e^{-\gamma_1(t-t')} \textrm{Erf}\left( \frac{v_1\sqrt{t-t'})}{\sqrt{4 D_1}}\right)-\chi_1\sigma_2\gamma_2\int_0^tdt' e^{-\gamma_2(t-t')}\textrm{Erf}\left( \frac{v_2t'-v_1t+\Delta X}{\sqrt{4 D_2 (t-t')}}\right)
 \end{split}\\
 \begin{split}
&S_2-1-\chi_2=\\ 
&-\chi_2\sigma_1\gamma_1\int_0^tdt' e^{-\gamma_1(t-t')} \textrm{Erf}\left( \frac{v_1t'-v_2t-\Delta X}{\sqrt{4 D_1 (t-t')}}\right)+\sigma_2\gamma_2\int_0^tdt' e^{-\gamma_2(t-t')}\textrm{Erf}\left( \frac{v_2\sqrt{t-t'}}{\sqrt{4 D_2}}\right)
 \end{split}
 \end{gathered} \label{eq:IGEP1}
 \end{equation}
Further, we can rearrange:
\begin{equation}
\frac{v_it'-v_jt\pm\Delta X}{\sqrt{4 D_i (t-t')}}=-\left(\frac{v_i}{\sqrt{4D_i}}\sqrt{t-t'}+\frac{(v_j-v_i) t\mp \Delta X}{\sqrt{4D_i(t-t')}}\right)
\end{equation}
which allows us to rewrite \eqref{eq:IGEP1} as:
\begin{equation}
\begin{gathered}
S_1-1-\chi_1=\sigma_1\gamma_1 I_1\left(\gamma_1,\frac{v_1}{\sqrt{4D_1}},t\right)+\chi_1\sigma_2\gamma_2 I_3\left(\gamma_2,\frac{-\Delta vt-\Delta X}{\sqrt{4D_2}} , \frac{v_2}{\sqrt{4D_2}},t \right) \\
S_2-1-\chi_2=\chi_2\sigma_1\gamma_1 I_3\left(\gamma_1,\frac{\Delta vt+\Delta X}{\sqrt{4D_1}} , \frac{v_1}{\sqrt{4D_1}},t \right)+\sigma_2\gamma_2 I_1\left(\gamma_2,\frac{v_2}{\sqrt{4D_2}},t\right)
\end{gathered}
\end{equation}
As the next step, we have to use explicit forms of integrals \eqref{eq:I1} and \eqref{eq:I3} and take the $t\to +\infty$ limit. For $I_1$ the result is analogous to the one-component case. To obtain the $t\to+\infty$ limit of $I_3$-dependent one must consider numerous limits. Most of them are elementary, but one needs to be considered with additional caution. Let us assume $a$, $b$ and $c$ are positive constant and consider the auxiliary problem:
\begin{equation}
\begin{split}
&\lim_{t\to+\infty} e^{2c tb+2|c|t\sqrt{\gamma_i+b^2}}\textrm{Erfc}\left(|c|\sqrt{t}+\sqrt{(\gamma_i+b^2)t} \right)=\lim_{t\to+\infty}\frac{\textrm{Erfc}\left(\left(|c|+\sqrt{\gamma_i+b^2}\right)\sqrt{t} \right)}{e^{-(2c b+2|c|\sqrt{\gamma_i+b^2})t}}\overset{L'H}{\propto}\\
&\overset{L'H}{\propto}\lim_{t\to+\infty}\frac{1}{2\sqrt{t}} \frac{e^{-\left(|c|+\sqrt{\gamma_i+b^2}\right)^2t}}{e^{-(2cb+2|c|\sqrt{\gamma_i+b^2})t}}=\lim_{t\to+\infty}\frac{1}{2\sqrt{t}}e^{-(|c|^2+\gamma_i+b^2-2cb)t}=\lim_{t\to+\infty}\frac{1}{2\sqrt{t}}e^{-(\gamma_i+(b-c)^2)t}=0
\end{split}
\end{equation}
where we employed L'Hospital's (L'H) rule. With this result at our disposal, we obtain:
\begin{equation}
\lim_{t\to+\infty}I_3\left(\gamma_1,\frac{\pm\Delta vt\mp\Delta X}{\sqrt{4D_1}} , \frac{v_1}{\sqrt{4D_1}},t \right)=\pm \frac{\textrm{sgn}(\Delta v)}{\gamma_i}
\end{equation}
and eventually:
\begin{equation}
\begin{gathered}
S_1-1-\chi_1=\sigma_1\frac{v_1}{\sqrt{4D_1\gamma_1+v_1^2}}-\chi_1\sigma_2\textrm{sgn}(\Delta v) \\
S_2-1-\chi_2=\chi_2\sigma_1\textrm{sgn}(\Delta v)+\sigma_2\frac{v_2}{\sqrt{4D_2\gamma_2+v_2^2}}
\end{gathered}
\end{equation}
We can rearrange these equations into the form:
\begin{equation}
\begin{gathered}
S_1-\chi_1(1-\sigma_2\textrm{sgn}(\Delta v))-1=\sigma_1\frac{v_1}{\sqrt{4D_1\gamma_1+v_1^2}}\\
S_2-\chi_2(1+\sigma_1\textrm{sgn}(\Delta v))-1=\sigma_2\frac{v_2}{\sqrt{4D_2\gamma_2+v_2^2}}
\end{gathered}
\end{equation}
When assumptions of the inter-growing domains are applied (i.e. the domains must exapand, $\textrm{sgn}(v_i)=-\sigma_i$), one can notice that $1+(-1)^i\sigma_j\textrm{sgn}(\Delta v)=2$. This shows that dynamics in IGEP phase is very similar to the non-interacting case, but it is $S_i-2\chi_i$ that controls the velocity instead of $S_i$.

%=========PERTURBATIVE ANALYSIS============
\section{Constant-velocity trajectory as the attractor of dynamics}
\subsection{Perturbative analysis of the one-component system}
In the previous section we derived the stability conditions by assuming the system adopts the constant velocity dynamics in the long run. However, the arguments provided earlier justify only that once the front is moving on the straight line in the time-space, it will do so indefinitely, i.e. that the constant velocity is the `stable point' of the dynamics. Here, we will discuss the influence of small perturbations. 

Let us assume that the position of the front reads:
 \begin{equation}
 X_i(t)=vt+\tilde X_i+\xi_i(t) \label{eq:X_perturbed}
 \end{equation}
 where $\xi_i(t)$ is a small perturbation. In the long time regime, $t\gg \max(\gamma_1^{-1},\gamma_2^{-1})$, the front dynamics reads:
 \begin{equation}
 \begin{split}
&S_i-1=-\sigma_i \gamma_i \int_0^t dt' e^{-\gamma_i(t-t')}\textrm{Erf}\left(\frac{-v(t-t')+\xi_i(t')-\xi_i(t)}{\sqrt{4D(t-t')}} \right) \label{eq:stab1}
\end{split}
 \end{equation}
We can use Taylor expansion in the integral term:
\begin{equation}
\begin{split}
&\int_0^t dt' e^{-\gamma_i(t-t')}\textrm{Erf}\left(\frac{-v(t-t')+\xi_i(t')-\xi_i(t)}{\sqrt{4D(t-t')}} \right)\simeq\\
&\simeq -\int_0^t dt' e^{-\gamma_i(t-t')}\textrm{Erf}\left(v\sqrt{\frac{t-t'}{4D(t-t')}} \right)+\int_0^t dt' e^{-(\gamma_i+\frac{v^2}{4D})(t-t')}\left(\frac{\xi_i(t')-\xi_i(t)}{\sqrt{\pi D(t-t')}} \right)=\\
&=-\int_0^t dt' e^{-\gamma_i(t-t')}\textrm{Erf}\left(v\sqrt{\frac{t-t'}{4D(t-t')}} \right)-\xi_i(t)\frac{\textrm{Erf}\left(\sqrt{(\gamma_i+\frac{v^2}{4D})t} \right)}{\sqrt{D\gamma_i+v^2/4}}+\int_0^t dt' \frac{e^{-(\gamma_i+\frac{v^2}{4D})(t-t')}}{\sqrt{\pi D(t-t')}}\xi_i(t')
\end{split}
\end{equation}
Inserting this expansion into \eqref{eq:stab1} and rearranging, we obtain:
 \begin{equation}
 \begin{split}
&S_i-1-\sigma_i\gamma_i\int_0^tdt'e^{-\gamma_i t}\textrm{Erf}\left(v\sqrt{\frac{t-t'}{4D}} \right)=\\
&=\sigma_i\gamma_i \xi_i(t)\frac{\textrm{Erf}\left(\sqrt{(\gamma_i+\frac{v^2}{4D})t} \right)}{\sqrt{D\gamma_i+v^2/4}}-\sigma_i\gamma_i\int_0^t dt' \frac{e^{-(\gamma_i+\frac{v^2}{4D})(t-t')}}{\sqrt{\pi D(t-t')}}\xi_i(t')
\end{split}
 \end{equation}
For large $t$ limit, the left-hand side of this equation approaches 0 when $v$ satisfies \eqref{eq:v_raw} and the error-function term approaches 1. Thus, we are left with:
\begin{equation}
\begin{split}
0=\sigma_i\xi_i(t)\frac{\gamma_i}{\sqrt{D\gamma_i+v^2/4}}-\sigma_i \gamma_i\int_0^t dt' \frac{e^{-(\gamma_i+\frac{v^2}{4D})(t-t')}}{\sqrt{\pi D(t-t')}}\xi_i(t')
\end{split}
\end{equation} 
This is the homogeneous Volterra integral equation of the second kind, whose only solution is $\xi_i(t)=const.$ (one can obtain it e.g. via the Laplace transform method). This shows that the deviations from the constant velocity dynamics must diminish in time and therefore $\lim_{t\to+\infty} \frac{X_i(t)}{t}=v_i$ is the attractor of the dynamics. However, the perturbation is still able to shift the long-term position of the front. 

\subsection{Perturbative analysis of the two-component system}
A similar approach can be now employed to analyze the influence of perturbation in two-component system. In this case, when we employ the perturbed constant velocity ansatz \eqref{eq:X_perturbed} (with $v_1=v_2=v$) the system reads:
 \begin{equation}
 \begin{gathered}
 \begin{split}
S_1-1-\chi_1=& -\sigma_1\gamma_1\int_0^tdt' e^{-\gamma_1(t-t')} \textrm{Erf}\left( \frac{v(t'-t)+\xi_1(t')-\xi_1(t)}{\sqrt{4 D_1 (t-t')}}\right)+\\
&-\chi_1\sigma_2\gamma_2\int_0^tdt' e^{-\gamma_2(t-t')}\textrm{Erf}\left( \frac{v(t'-t)+\Delta X+\xi_2(t')-\xi_1(t)}{\sqrt{4 D_2 (t-t')}}\right)
 \end{split}\\
 \begin{split}
S_2-1-\chi_2=&-\chi_2\sigma_1\gamma_1\int_0^tdt' e^{-\gamma_1(t-t')} \textrm{Erf}\left( \frac{v(t'-t)-\Delta X+\xi_1(t')-\xi_2(t)}{\sqrt{4 D_1 (t-t')}}\right)-\\
&-\sigma_2\gamma_2\int_0^tdt' e^{-\gamma_2(t-t')}\textrm{Erf}\left( \frac{v(t'-t)+\xi_2(t')-\xi_2(t)}{\sqrt{4 D_2 (t-t')}}\right)
 \end{split}
 \end{gathered} 
 \end{equation}
Repeating the steps from the one-component case, i.e. applying the Taylor expansions around the stable solution, rearranging and taking the large $t$ limit, where the non-perturbative part nullifies, we arrive at the system:
\begin{equation}
\begin{gathered}
\begin{split}
0=&\xi_1(t)\frac{\sigma_1\gamma_1}{\sqrt{D_1\gamma_1+v^2/4}}-\sigma_1\gamma_1\int_0^t dt' \frac{e^{-(\gamma_1+\frac{v^2}{4D_1})(t-t')}}{\sqrt{\pi D_1(t-t')}}\xi_1(t')+\\
&\xi_1(t)\frac{\chi_1\sigma_2\gamma_2}{\sqrt{D_2\gamma_2+v^2/4}}-\chi_1\sigma_2\gamma_2\int_0^t dt' \frac{e^{-(\gamma_2+\frac{v^2}{4D_2})(t-t')}}{\sqrt{\pi D_2(t-t')}}\xi_2(t')
\end{split} \\
\begin{split}
0=&\xi_2(t)\frac{\sigma_2\gamma_2}{\sqrt{D_2\gamma_2+v^2/4}}-\sigma_2\gamma_2\int_0^t dt' \frac{e^{-(\gamma_2+\frac{v^2}{4D_2})(t-t')}}{\sqrt{\pi D_2(t-t')}}\xi_2(t')+\\
&\xi_2(t)\frac{\chi_2\sigma_1\gamma_1}{\sqrt{D_1\gamma_1+v^2/4}}-\chi_2\sigma_1\gamma_1\int_0^t dt' \frac{e^{-(\gamma_1+\frac{v^2}{4D_1})(t-t')}}{\sqrt{\pi D_1(t-t')}}\xi_1(t')
\end{split}
\end{gathered}
\end{equation} 
This is a set of coupled homogeneous Volterra equations of the second kind. In order to obtain its solution, we can apply the Laplace transform to this system. Then it turns into:
 \begin{equation}
\begin{gathered}
0=\left(\frac{\sigma_1\gamma_1}{\sqrt{D_1\gamma_1+v^2/4}}+\frac{\chi_1\sigma_2\gamma_2}{\sqrt{D_2\gamma_2+v^2/4}}-\sigma_1\gamma_1\hat K_1(s) \right)\hat \xi_1(s)-\chi_1\sigma_2\gamma_2\hat K_2(s)\hat \xi_2(s)\\
0=-\chi_2\sigma_1\gamma_1 \hat K_1(s)\hat \xi_1(s)+\left(\frac{\sigma_2\gamma_2}{\sqrt{D_2\gamma_2+v^2/4}}+\frac{\chi_2\sigma_1\gamma_1}{\sqrt{D_1\gamma_1+v^2/4}}-\sigma_2\gamma_2\hat K_2(s) \right)\hat \xi_2(s)
\end{gathered}\label{eq:stab2}
\end{equation} 
where:
\begin{equation}
\begin{gathered}
\hat \xi_i(s)=\int_0^{+\infty}dt e^{-st}\xi_i(t)\\
 \hat K_i(s)=\int_0^{+\infty}dt \frac{e^{-st-(\gamma_i+\frac{v^2}{4D_i})t}}{\sqrt{\pi D_i t}}=\frac{1}{\sqrt{s+D_i\gamma_i+\frac{v^2}{4}}}
\end{gathered}
\end{equation}
The system of equations \eqref{eq:stab2} is linear and homogeneous, so it has no other solution than $\hat \xi_i(s)=0$ unless its determinant is 0. Let us investigate this possibility. For $s\neq 0$ the determinant, arranged by the powers of $\hat K_i(s)$, reads:
\begin{equation}
\begin{split}
&\det=\left(\frac{\sigma_1\gamma_1}{\sqrt{D_1\gamma_1+v^2/4}}+\frac{\chi_1\sigma_2\gamma_2}{\sqrt{D_2\gamma_2+v^2/4}}\right)\left(\frac{\sigma_2\gamma_2}{\sqrt{D_2\gamma_2+v^2/4}}+\frac{\chi_2\sigma_1\gamma_1}{\sqrt{D_1\gamma_1+v^2/4}} \right)\\
&-\sigma_1\gamma_1 \left(\frac{\sigma_2\gamma_2}{\sqrt{D_2\gamma_2+v^2/4}}+\frac{\chi_2\sigma_1\gamma_1}{\sqrt{D_1\gamma_1+v^2/4}} \right)\hat K_1(s)-\sigma_2\gamma_2 \left(\frac{\sigma_1\gamma_1}{\sqrt{D_1\gamma_1+v^2/4}}+\frac{\chi_1\sigma_2\gamma_2}{\sqrt{D_2\gamma_2+v^2/4}}\right)\hat K_2(s)\\
&+\gamma_1\gamma_2\sigma_1\sigma_2(1+\chi_1\chi_2)\hat K_1(s) \hat K_2(s)
\end{split}
\end{equation}
For $\det=0$ it must be satisfied that:
\begin{equation}
\begin{gathered}
\frac{\sigma_2\gamma_2}{\sqrt{D_2\gamma_2+v^2/4}}=-\frac{\chi_1^{-1}\sigma_1\gamma_1}{\sqrt{D_1\gamma_1+v^2/4}}\\
\frac{\sigma_2\gamma_2}{\sqrt{D_2\gamma_2+v^2/4}}=-\frac{\chi_2\sigma_1\gamma_1}{\sqrt{D_1\gamma_1+v^2/4}}\\
1+\chi_1\chi_2=0
\end{gathered}
\end{equation}
However, this system is contradictory (from the first two equations we get: $1-\chi_1\chi_2=0$), thus $\det$ must be non-zero.

The case $s=0$ corresponds to the constant solution $\xi_i(t)=\xi_i=const.$. Substituting the explicit form of $\hat K_i(0)$, equations \eqref{eq:stab2} are turned into:
\begin{equation}
\begin{gathered}
0=\frac{\chi_1\gamma_2}{\sqrt{D_2\gamma_2+v^2/4}}(\hat \xi_1(0)-\hat \xi_2(0))\\
0=-\frac{\chi_2\gamma_1}{\sqrt{D_1\gamma_1+v^2/4}}(\hat \xi_1(0)-\hat \xi_2(0))
\end{gathered}
\end{equation}
whose determinant is 0. Thus, indeed $\xi_i(t)=const.$ is the only possible solution. Once again it shows that any deviation from the constant velocity dynamics must diminish in the long run, i.e. indeed $\lim_{t\to+\infty}\frac{X_i(t)}{t}=v$ is the attractor of the dynamics. However, this result conveys that $\Delta X$ is also the attractor. The reason is that for a given set of system parameters the system \eqref{eq:main_fin} defines a unique pair of $v$ and $\Delta X$. Thus, once $v$ is reestablished, so must be $\Delta X$. In other words, the contact zone between the domains can shift as a whole due to the perturbations, but the system will tend to restore the inter-domain distance $\Delta X$.

%===========FINITE SIZE EFFECTS===============
\section{Effects of finite system size $L$}
In general, the influence of the finite-size on the behavior of quasi-non-linear model is a complex problem, whose full discussion is beyond the scope of this supplement. However, a few simpler results can be provided to show that the dynamics of systems with $L\gg\lambda_i$ is well approximated by the predictions for infinite systems.

\subsection{$G_i(\Delta x,\Delta t)$ in finite-size systems}
When the differential equations \eqref{eq:main} are solved in the finite-$L$ system, the integral kernel in solution \eqref{eq:solution} is not a closed-form function, but it is given by a Fourier series, whose detailed form depends on the boundary conditions. For example, for reflective boundaries, function $G_i(\Delta x,\Delta t)$ reads:
\begin{equation}
G_i(\Delta x, \Delta t)=\frac{1}{L} e^{-\gamma_i \Delta t}\sum_{k=-\infty}^{+\infty} e^{-\frac{(\pi k)^2}{L^2}D_i \Delta t}\cos \frac{\pi k \Delta x}{L}
\end{equation}
For short $\Delta t$ this function closely resembles a Gaussian, but deviates from it for larger $\Delta t$. In order to quantify this effect, let us notice that due to the exponential decay in the prefactor, the significant values of $G_i(\Delta x,\Delta t)$ are concentrated for $0<\Delta t<\gamma_i^{-1}$. At this time-scale, the effective width of the Gaussian is of the order of $\sqrt{D_i/\gamma_i}=\lambda_i$. Thus, for systems where $L\gg \lambda_i$ the deviations of $G_i(\Delta x,\Delta t)$ from Gaussianity are minimal.

\subsection{Constant velocity ansatz in finite-size systems: TGEP vs. SGEP} 
The constant velocity ansatz is exact in infinite systems ($L\to+\infty$) and for $t\to+\infty$. However, even in infinite systems the constant-velocity dynamics is reached only asymptotically in time. For this reason, the constant velocity ansatz is necessarily only the approximation in finite-$L$ systems, as traveling fronts always reach the system boundary in finite time. Thus, the analysis of TGEP dynamics in finite-$L$ systems must be done in the transient regime and is beyond the scope of this supplement. 

The situation is much simplified in the case of stabilized fronts (i.e. SGEPs). In this case the fronts eventually stop and cannot reach the system boundaries even for $t\to+\infty$. 
A separate problem, which we are not able to address here, is for which initial conditions SGEPs can be achieved. However, assuming that such initial states exist, we can repeat the derivation of stability conditions in the finite-$L$ case. In the course of this derivation we must also explicitly assume that $X_i(t)=vt+\tilde X_i$ with $v=0$.  Eventually, taking the $t\to+\infty$ limit and assuming geometry $\sigma_1=-1$ and $\sigma_2=1$, we arrive at the expressions:
\begin{equation}
\begin{gathered}
S_1=\left(1-e^{-\frac{L/2+\tilde X_1}{\lambda_1}} \right)+\chi_1\left(1-e^{-\frac{L/2-\tilde X_1}{\lambda_2}}-\textrm{sgn}(\tilde X_2-\tilde X_1)\left( 1-e^{-\frac{|\tilde X_2-\tilde X_1|}{\lambda_2}}  \right)\right)\\
S_2=\chi_2\left(1-e^{-\frac{L/2+\tilde X_2}{\lambda_1}}-\textrm{sgn}(\tilde X_2-\tilde X_1)\left( 1-e^{-\frac{|\tilde X_2-\tilde X_1|}{\lambda_1}}  \right)\right)+\left(1-e^{-\frac{L/2-\tilde X_2}{\lambda_2}} \right)
\end{gathered}\label{eq:finite_L_stab}
\end{equation}
This system of equations is identical to system \eqref{eq:main_stable}, up to the presence of additional exponential terms $e^{-\frac{L/2\pm \tilde X_i}{\lambda_i}}$. It is via these terms that the influence of boundaries manifests in the stabilization conditions.  It is straightforward to notice that  when $L/2-|\tilde X_i|\gg \lambda_i$ (which implies  $L\gg\lambda_i$), these terms are effectively close to 0 and the stabilization conditions for the infinite system are virtually identical to the `in the bulk' stabilization conditions for finite-$L$ systems. Significant deviations occur only for $\tilde X_i$ situated within the distance $\lambda_i$ from the system boundary, where these exponential terms have relatively large magnitude. 

The important difference between system \eqref{eq:main_stable} and \eqref{eq:finite_L_stab} is that the former is an overdefined system, which allows us to find only $\Delta X$, while the latter defines two absolute positions of the fronts. In a certain way, the information about absolute positions of the fronts is lost in the $L\to+\infty$ case, while retained in the finite-$L$ case. Nevertheless, extracting this information from the system \eqref{eq:finite_L_stab} is difficult, as for $L\gg\lambda_i$ its numerical solutions seem highly unstable. This is caused by the fact that additional exponential terms in this limit are very small, so numerically, the system behaves as over-defined. 

\subsection{The influence of boundary vicinity}
The final group of effects is related to the influence of boundary vicinity on the trajectory of $X_i(t)$. Let us start with one-component system and  write the analog of expression \eqref{eq:one_comp} for the finite-$L$ system:
\begin{equation}
\frac{C_i}{\epsilon_{ii}H_i}=\sigma_i\int_0^t dt'\int^{\sigma_i L/2}_{X_i(t')}dx' G_i(x'-X_i(t),t-t')\label{eq:one_finiteL}
\end{equation}
Looking at Fig. \ref{fig:int1} one can notice that it should be discerned between $X_i(t)$ approaching the non-activated side of the system and $X_i(t)$ retracting towards the the activated side. In the former case there is no influence of boundary as it is not able to modify the integration region in \eqref{eq:one_finiteL}. In the latter case, the influence is significant as soon as $X_i(t)$ is within the $\simeq\lambda_i$ distance from the boundary (we already estimated that the effective width of $G_i(\Delta X,\Delta t)$ is of the $\lambda_i$ order). Since the presence of the boundary restricts the area of integration, the trajectory must compensate this effect to maintain the constant value of the integral. This manifests as rapid `attraction' toward the boundary.

In case of TGEP (in two-component systems), the influence of boundaries is observed no matter which side of the system is approached. Fig. \ref{fig:int2} shows that $X_i(t)$ dynamics is governed by the integrals taken towards both ends of the system. Thus, as the fronts approach either system boundary, at least two of these integrals are disturbed. This usually manifests as the rapid attraction of the fronts towards the approached boundary, when the fronts get withing adequate $\lambda_i$ distance from it.

%===========EXISTENCE OF SOLUTIONS==================
\section{Analytical estimation of the minimal region where the system defining $v$ and $\Delta X$ has no solutions}
We assume $\sigma_1=-1$ and $\sigma_2=1$. Let us express the main set of equations for long-term behavior, \eqref{eq:main_fin} as:
 \begin{equation}
 \begin{gathered}
 S_1=1-V_1+\chi_1\left(1-\textrm{sgn}(\Delta X)+e^{\frac{v\Delta X}{2D_2}\left(1-\frac{\textrm{sgn}(\Delta X)}{V_2} \right)}\left( V_2+\textrm{sgn}(\Delta X)\right) \right) \\
 S_2=1+V_2+\chi_2\left(1-\textrm{sgn}(\Delta X)-e^{-\frac{v\Delta X}{2D_1}\left(1+\frac{\textrm{sgn}(\Delta X)}{V_1} \right)}\left( V_1-\textrm{sgn}(\Delta X)\right) \right)
 \end{gathered}
 \end{equation}
 where:
 \begin{align}
 V_i=\frac{v}{\sqrt{4D_i\gamma_i+v^2}}
 \end{align}
 Importantly, $V_i$ takes values between -1 and 1. Both $V_i$ have also the same sign. We can rearrange the main equations to obtain:
  \begin{equation}
 \begin{gathered}
 \frac{S_1-1+V_1-\chi_1+\chi_1\textrm{sgn}(\Delta X)}{\chi_1\left( V_2+\textrm{sgn}(\Delta X)\right)}=e^{\frac{v\Delta X}{2D_2}\left(1-\frac{\textrm{sgn}(\Delta X)}{V_2} \right)} \\
\frac{ S_2-1-V_2-\chi_2+\chi_2 \textrm{sgn}(\Delta X)}{\chi_2\left(\textrm{sgn}(\Delta X)-V_1\right)}=e^{-\frac{v\Delta X}{2D_1}\left(1+\frac{\textrm{sgn}(\Delta X)}{V_1} \right)}
 \end{gathered}
 \end{equation}
 We will now consider 4 combinations of $v$ and $\Delta X$ being greater or smaller than 0 and find the minimal region of $(\chi_1,\chi_2)$ plane, where \eqref{eq:main_fin} has solutions. In this section we limit our considerations to $\chi_i<0$ and $0<S_i<1$.
 
\begin{figure}[h]
\centering
\includegraphics[width=0.5\textwidth]{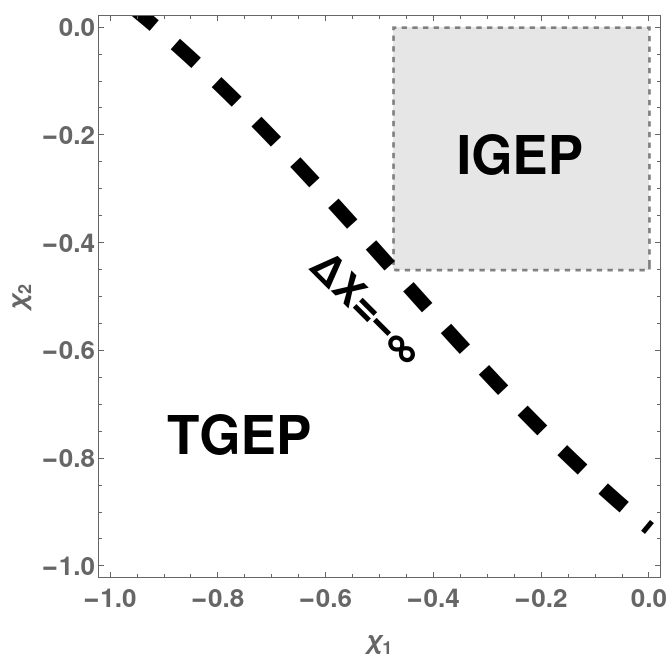}
\caption{Grayed-out square: graphical illustration of \eqref{eq:no_sol}, i.e. the analytically found minimal region where Eqs. (11) from the main text have no solution for $(v,\Delta X)$. \label{fig:region_min}}
\end{figure}

\begin{itemize}
\item $v>0,\Delta X>0$: In this case, we obtain:
\begin{equation}
\begin{gathered}
 \frac{S_1-1+|V_1|}{\chi_1\left( |V_2|+1\right)}=e^{\frac{|v\Delta X|}{2D_2}\left(1-\frac{1}{|V_2|} \right)} \\
\frac{ S_2-1-|V_2|}{\chi_2\left(1-|V_1|\right)}=e^{-\frac{|v\Delta X|}{2D_1}\left(1+\frac{1}{|V_1|} \right)}
 \end{gathered}
 \end{equation}
 From analyzing the right-hand of this equation, we conclude the inequalities:
\begin{equation}
\begin{gathered}
1> \frac{S_1-1+|V_1|}{\chi_1\left( |V_2|+1\right)}>0 \\
1>\frac{ S_2-1-|V_2|}{\chi_2\left(1-|V_1|\right)}>0
 \end{gathered}
 \end{equation}
After rearranging:
 \begin{equation}
 \begin{aligned}
 \chi_1<\frac{S_1-1+|V_1|}{|V_2|+1}&,&|V_1|<1-S_1 \\
 \chi_2<\frac{S_2-1-|V_2|}{1-|V_1|}&,&S_2-1<|V_2|
 \end{aligned}
 \end{equation}
Taking into account the range of values for $V_i$ and $S_i$, this further is extremized by:
\begin{equation}
\begin{gathered}
\chi_1<0\\
\chi_2<S_2-1
 \end{gathered}
 \end{equation}
 
 \item $v<0,\Delta X<0$:
 \begin{equation}
\begin{gathered}
 \frac{S_1-1-|V_1|-2\chi_1}{\chi_1\left( -|V_2|-1\right)}=e^{\frac{|v\Delta X|}{2D_2}\left(1-\frac{1}{|V_2|} \right)} \\
\frac{ S_2-1+|V_2|-2\chi_2}{\chi_2\left(-1+|V_1|\right)}=e^{-\frac{|v\Delta X|}{2D_1}\left(1+\frac{1}{|V_1|} \right)}
 \end{gathered}
 \end{equation}
 This results in inequalities:
 \begin{equation}
\begin{gathered}
1> \frac{S_1-1-|V_1|-2\chi_1}{\chi_1\left( -|V_2|-1\right)}>0 \\
1>\frac{ S_2-1+|V_2|-2\chi_2}{\chi_2\left(-1+|V_1|\right)}>0
 \end{gathered}
 \end{equation}
  After rearranging:
\begin{equation}
 \begin{aligned}
 \chi_1>\frac{S_1-1-|V_1|}{1-|V_2|}&,&\frac{S_1-1-|V_1|}{2}>\chi_1 \\
 \chi_2>\frac{S_2-1+|V_2|}{1+|V_1|}&,&\frac{S_2-1+|V_2|}{2}>\chi_2
 \end{aligned}
 \end{equation}
 Extremizing these inequalities, we get:
\begin{equation}
\begin{gathered}
\frac{S_1-1}{2}>\chi_1\\
\frac{S_2}{2}>\chi_2>S_2-1
 \end{gathered}
 \end{equation}
 
 \item $v>0,\Delta X<0$:
 \begin{equation}
\begin{gathered}
 \frac{S_1-1+|V_1|-2\chi_1}{\chi_1\left(|V_2|-1\right)}=e^{-\frac{|v\Delta X|}{2D_2}\left(1+\frac{1}{|V_2|} \right)} \\
\frac{ S_2-1-|V_2|-2\chi_2}{\chi_2\left(-1-|V_1|\right)}=e^{\frac{|v\Delta X|}{2D_1}\left(1-\frac{1}{|V_1|} \right)}
 \end{gathered}
 \end{equation}
 This results in inequalities:
 \begin{equation}
\begin{gathered}
1>\frac{S_1-1+|V_1|-2\chi_1}{\chi_1\left(|V_2|-1\right)}>0 \\
1>\frac{ S_2-1-|V_2|-2\chi_2}{\chi_2\left(-1-|V_1|\right)}>0
 \end{gathered}
 \end{equation}
 After rearranging:
  \begin{equation}
 \begin{aligned}
 \chi_1>\frac{S_1-1+|V_1|}{1+|V_2|}&,&\frac{S_1-1+|V_1|}{2}>\chi_1 \\
 \chi_2>\frac{S_2-1-|V_2|}{1-|V_1|}&,&\frac{S_2-1-|V_2|}{2}>\chi_2
 \end{aligned}
 \end{equation}
 Extremizing these inequalities, we get:
 \begin{equation}
\begin{gathered}
\frac{S_1}{2}>\chi_1>S_1-1\\
\frac{S_2-1}{2}>\chi_2>-\infty
 \end{gathered}
 \end{equation}
 
 \item $v<0,\Delta X>0$:
 
 \begin{equation}
\begin{gathered}
 \frac{S_1-1-|V_1|}{\chi_1\left(1-|V_2|\right)}=e^{-\frac{|v\Delta X|}{2D_2}\left(1+\frac{1}{|V_2|} \right)} \\
\frac{ S_2-1+|V_2|}{\chi_2\left(1+|V_1|\right)}=e^{\frac{|v\Delta X|}{2D_1}\left(1-\frac{1}{|V_1|} \right)}
 \end{gathered}
 \end{equation}
 This results in inequalities:
 \begin{equation}
\begin{gathered}
1> \frac{S_1-1-|V_1|}{\chi_1\left(1-|V_2|\right)}>0 \\
1>\frac{ S_2-1+|V_2|}{\chi_2\left(1+|V_1|\right)}>0
 \end{gathered}
 \end{equation}
After rearranging:
 \begin{equation}
 \begin{aligned}
 \chi_1<\frac{S_1-1-|V_1|}{1-|V_2|}&,&S_1-1<|V_1| \\
 \chi_2<\frac{S_2-1+|V_2|}{1+|V_1|}&,&|V_2|<1-S_2
 \end{aligned}
 \end{equation}
 Substituting in the upper $|V_1|=0$ and $|V_2|=0$ and in the lower one $|V_2|=1-S_2$:
 \begin{equation}
\begin{gathered}
\chi_1<S_1-1\\
\chi_2<0
 \end{gathered}
 \end{equation}
  \end{itemize}
 Summarizing these considerations, the minimal boundaries for a region where solutions do not exist read (see Fig.~\ref{fig:region_min}):
 \begin{align}
 \chi_i>\frac{S_i-1}{2} \label{eq:no_sol}
 \end{align}

 \section{Analytical prediction of the $\Delta X=-\infty$ line on the $(\chi_1,\chi_2)$ plot} 
 We can also localize the line of $\Delta X=-\infty$ as a possible approximation for the boundary region where solutions exist. By taking this limit (for $\sigma_1=-1$ and $\sigma_2=1$), the system reduces to:
 \begin{equation}
 \begin{gathered}
 S_1=1-\frac{v}{\sqrt{4D_1\gamma_1+v^2}}+2\chi_1\\
 S_2=1-\frac{v}{\sqrt{4D_2\gamma_2+v^2}}+2\chi_2
 \end{gathered}
 \end{equation} 
Solving for $v$ we obtain:
 \begin{equation}
 \begin{gathered}
\frac{v^2}{4D_1\gamma_1+v^2}=(1-S_1+2\chi_1)^2\\
\frac{v^2}{4D_2\gamma_2+v^2} =(1-S_2+2\chi_2)^2
 \end{gathered}
 \end{equation}
 and eventually:
 \begin{equation}
 \begin{gathered}
v^2=4D_1\gamma_1\frac{(1-S_1+2\chi_1)^2}{1-(1-S_1+2\chi_1)^2}\\
v^2 =4D_2\gamma_2\frac{(1-S_2+2\chi_2)^2}{1-(1-S_2+2\chi_21)^2}
 \end{gathered}
 \end{equation}
Finally, we must demand that both $v^2$ are equal to each other, as the system is overdefined, which defines the implicit shape on the $(\chi_1,\chi_2)$ plane:
\begin{equation}
4D_1\gamma_1\frac{(1-S_1+2\chi_1)^2}{1-(1-S_1+2\chi_1)^2}=4D_2\gamma_2\frac{(1-S_2+2\chi_2)^2}{1-(1-S_2+2\chi_2)^2}
\end{equation}
For practical reasons we can calculate $\chi_2$ as the function of $\chi_1$:
\begin{equation}
\chi_2=\frac{1}{2}\left( S_2-1+\frac{\textrm{sgn}(v)}{\sqrt{1+\frac{D_2\gamma_2}{D_1\gamma_1}\frac{1-(1-S_1+2\chi_1)^2}{(1-S_1+2\chi_1)^2}}}\right)
\end{equation}

%=========INTEGRALS=============
\section{Important integrals used in the derivation}
\subsection{Type 1}
There are three types of integrals that we need to calculate. It is feasible to focus on the indefinite versions. The simplest type reads:
 \begin{equation}
 \begin{split}
\int dt e^{-\gamma t} \textrm{Erf}(a\sqrt{t})=-\frac{e^{-\gamma t}}{\gamma}\textrm{Erf}(a\sqrt{t})+\frac{a}{\gamma}\int dt \frac{e^{-\gamma t-a^2 t}}{\sqrt{\pi t}}=-\frac{e^{-\gamma t}}{\gamma}\textrm{Erf}(a\sqrt{t})+\frac{a}{\gamma}\frac{\textrm{Erf}\left(\sqrt{(a^2+\gamma)t} \right)}{\sqrt{(a^2+\gamma)}}
\end{split}
 \end{equation}
\subsection{Type 2}
Another integral is more demanding and it reads:
 \begin{equation}
 \begin{split}
\int dt e^{-\gamma t}\textrm{Erf}(a/\sqrt{t})&=-\frac{ e^{-\gamma t}}{\gamma}\textrm{Erf}(a/\sqrt{t}) -\frac{a}{\gamma \sqrt{\pi}}\int dt \frac{e^{-\gamma t-\frac{a^2}{t}}}{t^{3/2}}
\end{split}
 \end{equation}
where in the first step we integrated by parts. Further, we make the substitution: $t=\frac{|a|}{\sqrt{\gamma}}\frac{1}{x^2}$ so $dt=-2\frac{|a|}{\sqrt{\gamma}}\frac{dx}{x^3}$ and $\lambda=|a|\sqrt{\gamma}$. It is important to comment that instead of $|a|$ we could use $a$ or $-a$, but these choices lead to different final results for $a<0$ (for $a$ we eventually encounter divergence for negative front velocity). Under our substitution, the integral reads: 
 \begin{equation}
\int dt e^{-\gamma t}\textrm{Erf}(a/\sqrt{t})=-\frac{ e^{-\gamma t}}{\gamma}\textrm{Erf}(a/\sqrt{t}) +\textrm{sgn}(a)\frac{2}{\gamma}\sqrt{\frac{\lambda}{ \pi }}\int dx e^{-\lambda(x^2+\frac{1}{x^2})}
 \end{equation}
Let us assign:
 \begin{equation}
I=\int dx e^{-\lambda(x^2+\frac{1}{x^2})}
 \end{equation}
 One can now notice that the expression in the exponent in $I$ can be represented in two equivalent ways:
 \begin{equation}
 x^2+\frac{1}{x^2}=(x+\frac{1}{x})^2-2=(x-\frac{1}{x})^2+2
 \end{equation}
Thus, we can rewrite the integral in the following manner:
 \begin{equation}
 \begin{split}
I&= \int dx e^{-\lambda(x^2+\frac{1}{x^2})}\left(\frac{1}{2}+\frac{1}{2}+\frac{1}{2x^2}-\frac{1}{2x^2} \right)=\\
 &= \int dx \left(\frac{e^{-2\lambda-\lambda\left(x-\frac{1}{x} \right)^2}}{2}+\frac{e^{2\lambda-\lambda\left(x+\frac{1}{x} \right)^2}}{2}+\frac{e^{-2\lambda-\lambda\left(x-\frac{1}{x} \right)^2}}{2x^2}-\frac{e^{2\lambda-\lambda\left(x+\frac{1}{x} \right)^2}}{2x^2} \right)=\\
 &=\frac{1}{2}\int dx \left(e^{-2\lambda-\lambda\left(x-\frac{1}{x} \right)^2}(1+\frac{1}{x^2})+e^{2\lambda-\lambda\left(x+\frac{1}{x} \right)^2}(1-\frac{1}{x^2}) \right)
 \end{split}
 \end{equation}
Now we can make the substitutions: $u_{\pm}=x\pm\frac{1}{x}$ so $du_{\pm}=(1\mp\frac{1}{x^2})dx$, i.e.:
 \begin{equation}
 \begin{split}
 I&=\frac{e^{-2\lambda}}{2}\int du_- e^{-\lambda u_-^2}+\frac{e^{2\lambda}}{2}\int du_+ e^{-\lambda u_+^2}=\\
 &=\frac{1}{4}\sqrt{\frac{\pi}{\lambda}} \left( e^{-2\lambda}\textrm{Erf}(\sqrt{\lambda}u_-)+e^{2\lambda}\textrm{Erf}(\sqrt{\lambda}u_+) \right)=\\
 &=\frac{1}{4}\sqrt{\frac{\pi}{\lambda}} \left[ e^{-2\lambda}\textrm{Erf}\left(\sqrt{\lambda}(x-\frac{1}{x})\right)+e^{2\lambda}\textrm{Erf}\left(\sqrt{\lambda}(x+\frac{1}{x})\right) \right]
 \end{split}
 \end{equation}
 Eventually, we obtain the result:
 \begin{equation}
 \int dt e^{-\gamma t}\textrm{Erf}(\frac{a}{\sqrt{t}})=-\frac{e^{-\gamma t}}{\gamma}\textrm{Erf}\left(\frac{a}{\sqrt{t}}\right)+\frac{\textrm{sgn}(a)}{2\gamma}\left[e^{-2|a|\sqrt{\gamma}}\textrm{Erf}\left(\frac{|a|}{\sqrt{t}}-\sqrt{\gamma t} \right)+e^{2|a|\sqrt{\gamma}}\textrm{Erf}\left(\frac{|a|}{\sqrt{t}}+\sqrt{\gamma t} \right)\right]
  \end{equation}

\subsection{Type 3}
 The last integral reads:
 \begin{equation}
 \int dt e^{-\gamma t}\textrm{Erf}\left(b\sqrt{t}+\frac{a}{\sqrt{t}} \right)
 \end{equation}
 Integrating by parts, we obtain:
  \begin{equation}
  \begin{split}
&\int dt e^{-\gamma t}\textrm{Erf}\left(b\sqrt{t}+\frac{a}{\sqrt{t}} \right)=-\frac{e^{-\gamma t}}{\gamma}\textrm{Erf}\left(b\sqrt{t}+\frac{a}{\sqrt{t}} \right)+2\int dt \frac{e^{-\gamma t-b^2t-2ab-a^2/t}}{\gamma\sqrt{\pi}}\left(\frac{b}{2\sqrt{t}}-\frac{a}{2t^{3/2}} \right)=\\
&=-\frac{e^{-\gamma t}}{\gamma}\textrm{Erf}\left(b\sqrt{t}+\frac{a}{\sqrt{t}} \right)+\frac{be^{-2ab}}{\gamma\sqrt{\pi}} \int dt \frac{e^{-(\gamma+b^2)t-\frac{a^2}{t}}}{\sqrt{t}}-\frac{ae^{-2ab}}{\gamma\sqrt{\pi}}\int dt \frac{e^{-(\gamma+b^2)t-\frac{a^2}{t}}}{t^{3/2}}
 \end{split}
 \end{equation}
 The $a$-multiplied term has already appeared in the previous integral, albeit with slightly different constants. Thus we must focus on the $b$-multiplied term, which, under the change of variables $t=\frac{|a|}{\sqrt{\gamma+b^2}}x^2$ (so $2\frac{\sqrt{|a|}}{(\gamma+b^2)^{1/4}}dx=\frac{dt}{\sqrt{t}}$), reads:
 \begin{equation}
 \int dt \frac{e^{-(\gamma+b^2)t-\frac{a^2}{t}}}{\sqrt{t}}=2\frac{\sqrt{|a|}}{(\gamma+b^2)^{1/4}}\int dx e^{-|a|\sqrt{\gamma+b^2}(x^2+\frac{1}{x^2})}
 \end{equation}
 This integral has also already been calculated. Thus, we can provide the final result instantly:
 \begin{equation}
 \begin{split}
 &\int dt e^{-\gamma t}\textrm{Erf}\left(b\sqrt{t}+\frac{a}{\sqrt{t}} \right)=-\frac{e^{-\gamma t}}{\gamma}\textrm{Erf}\left(b\sqrt{t}+\frac{a}{\sqrt{t}} \right)+\\
 &+\frac{be^{-2ab}}{2\gamma\sqrt{\gamma+b^2}}  \left[ e^{-2|a|\sqrt{\gamma+b^2}}\textrm{Erf}\left(\sqrt{(\gamma+b^2)t}-\frac{|a|}{\sqrt{t}}\right)+e^{2|a|\sqrt{\gamma+b^2}}\textrm{Erf}\left(\sqrt{(\gamma+b^2)t}+\frac{|a|}{\sqrt{t}}\right) \right]+\\
 &+\textrm{sgn}(a)\frac{e^{-2ab}}{2\gamma}\left[e^{-2|a|\sqrt{\gamma+b^2}}\textrm{Erf}\left(\frac{|a|}{\sqrt{t}}-\sqrt{(\gamma +b^2)t} \right)+e^{2|a|\sqrt{\gamma+b^2}}\textrm{Erf}\left(\frac{|a|}{\sqrt{t}}+\sqrt{(\gamma+b^2 )t} \right)\right]=\\
=&-\frac{e^{-\gamma t}}{\gamma}\textrm{Erf}\left(b\sqrt{t}+\frac{a}{\sqrt{t}} \right)-\frac{e^{-2ab}}{2\gamma}  \left[ \left(\frac{b}{\sqrt{\gamma+b^2}}-\textrm{sgn}(a) \right)e^{-2|a|\sqrt{\gamma+b^2}}\textrm{Erf}\left(\frac{|a|}{\sqrt{t}}-\sqrt{(\gamma+b^2)t}\right)+\right.\\
 &\left.+\left(\frac{b}{\sqrt{\gamma+b^2}}+\textrm{sgn}(a) \right)e^{2|a|\sqrt{\gamma+b^2}}\textrm{Erf}\left(\sqrt{(\gamma+b^2)t}+\frac{|a|}{\sqrt{t}})\right) \right]
 \end{split}
 \end{equation}

 \subsection{Definite integrals}
  Finally, we can provide the definite integrals:
 \begin{equation}
I_1(\gamma_i,b,t)= \int_0^t dt' e^{-\gamma t'} \textrm{Erf}(b\sqrt{t'})=-\frac{e^{-\gamma t}}{\gamma}\textrm{Erf}(b\sqrt{t})+\frac{b}{\gamma}\frac{\textrm{Erf}\left(\sqrt{(b^2+\gamma)t} \right)}{\sqrt{(b^2+\gamma)}} \label{eq:I1}
 \end{equation}
 \begin{equation}
 \begin{split}
 I_2(\gamma_i,a,t)=&\int_0^t dt' e^{-\gamma t'}\textrm{Erf}(\frac{a}{\sqrt{t'}})=\frac{\textrm{sgn}(a)}{\gamma}-\frac{e^{-\gamma t}}{\gamma}\textrm{Erf}\left(\frac{a}{\sqrt{t}}\right)-\\
 &-\frac{\textrm{sgn}(a)}{2\gamma}\left[e^{-2|a|\sqrt{\gamma}}\textrm{Erfc}\left(\frac{|a|}{\sqrt{t}}-\sqrt{\gamma t} \right)+e^{2|a|\sqrt{\gamma}}\textrm{Erfc}\left(\frac{|a|}{\sqrt{t}}+\sqrt{\gamma t} \right)\right]
 \end{split}\label{eq:I2}
\end{equation} 
\begin{equation}
 \begin{split}
&I_3(\gamma_i,a,b,t)=\int_0^t dt' e^{-\gamma t'}\textrm{Erf}\left(b\sqrt{t'}+\frac{a}{\sqrt{t'}} \right)=\frac{\textrm{sgn}(a)}{\gamma}-\frac{e^{-\gamma t}}{\gamma}\textrm{Erf}\left(b\sqrt{t}+\frac{a}{\sqrt{t}} \right)+\\
 &+\frac{e^{-2ab}}{2\gamma}  \left[ \left(\frac{b}{\sqrt{\gamma+b^2}}-\textrm{sgn}(a) \right)e^{-2|a|\sqrt{\gamma+b^2}}\textrm{Erfc}\left(\frac{|a|}{\sqrt{t}}-\sqrt{(\gamma+b^2)t}\right)-\right.\\
 &\left.-\left(\frac{b}{\sqrt{\gamma+b^2}}+\textrm{sgn}(a) \right)e^{2|a|\sqrt{\gamma+b^2}}\textrm{Erfc}\left(\sqrt{(\gamma+b^2)t}+\frac{|a|}{\sqrt{t}})\right) \right]
 \end{split} \label{eq:I3}
\end{equation}